\documentclass[10pt, conference]{IEEEtran}
\IEEEoverridecommandlockouts
\usepackage{cite}
\usepackage{amsmath,amssymb,amsfonts}
\usepackage{algorithmic}
\usepackage{graphicx}
\usepackage{textcomp}
\usepackage{xcolor}
\usepackage{collab}
\usepackage{url}
\usepackage{subfigure}
\usepackage{booktabs}
\usepackage{threeparttable}
\usepackage{multirow}
\usepackage{verbatim}
\usepackage{balance}

\newcommand{\wasm}{Wasm\xspace}
\newcommand{\ora}{\textit{oracle ratio}\xspace}
\newcommand{\dd}{\textit{deviation degree}\xspace}
\newcommand{\ics}{\textit{issue-related code snippet}\xspace}
\newcommand{\wasmdiff}{\textit{WarpDiff}\xspace}

\def\BibTeX{{\rm B\kern-.05em{\sc i\kern-.025em b}\kern-.08em
    T\kern-.1667em\lower.7ex\hbox{E}\kern-.125emX}}

\begin{document}

\title{Revealing Performance Issues in Server-side WebAssembly Runtimes via Differential Testing
}

\author{
\IEEEauthorblockN{Shuyao Jiang$^*$, Ruiying Zeng$^{\dag\ddag}$, Zihao Rao$^{\dag\ddag}$, Jiazhen Gu$^*$, Yangfan Zhou$^{\dag\ddag}$, and Michael R. Lyu$^*$}
\IEEEauthorblockA{\textit{$^*$ Department of Computer Science and Engineering, The Chinese University of Hong Kong, Hong Kong, China}}
\IEEEauthorblockA{\textit{$^\dag$ School of Computer Science, Fudan University, Shanghai, China}}
\IEEEauthorblockA{\textit{$^\ddag$ Shanghai Key Laboratory of Intelligent Information Processing, Shanghai, China}}
\IEEEauthorblockA{syjiang21@cse.cuhk.edu.hk, \{ryzeng22, zhrao23\}@m.fudan.edu.cn, \\
jiazhengu@cuhk.edu.hk, zyf@fudan.edu.cn, lyu@cse.cuhk.edu.hk}
}

\maketitle

\begin{abstract}
WebAssembly (Wasm) is a bytecode format originally serving as a compilation target for Web applications. It has recently been used increasingly on the server side, \textit{e.g.}, providing a safer, faster, and more portable alternative to Linux containers. With the popularity of server-side Wasm applications, it is essential to study performance issues (\textit{i.e.}, abnormal latency) in Wasm runtimes, as they may cause a significant impact on server-side applications. However, there is still a lack of attention to performance issues in server-side Wasm runtimes. In this paper, we design a novel differential testing approach \textit{WarpDiff} to identify performance issues in server-side Wasm runtimes. The key insight is that in normal cases, the execution time of the same test case on different Wasm runtimes should follow an \textit{oracle ratio}. We identify abnormal cases where the execution time ratio significantly deviates from the \textit{oracle ratio} and subsequently locate the Wasm runtimes that cause the performance issues. We apply \textit{WarpDiff} to test five popular server-side Wasm runtimes using 123 test cases from the LLVM test suite and demonstrate the top 10 abnormal cases we identified. We further conduct an in-depth analysis of these abnormal cases and summarize seven performance issues, all of which have been confirmed by the developers. We hope our work can inspire future investigation on improving Wasm runtime implementation and thus promoting the development of server-side Wasm applications.
\end{abstract}


\section{Introduction}\label{sec:intro}

WebAssembly (abbreviated \wasm) is a static low-level bytecode format designed as a portable compilation target for the Web~\cite{haas2017bringing,reiser2017accelerate,wagner2017webassembly}.
\wasm bytecodes are fast to compile and run, portable across browsers and architectures, and provide guarantees of type and memory safety. 
Such characteristics make \wasm to be increasingly adopted outside the Web context.
In particular, \wasm has been considered a better isolation mechanism than containers in cloud environments~\cite{shillaker2020faasm,gackstatter2022pushing,eriksson2021containerizing,jain2022extending,long2020lightweight}, 
since it provides a higher level of abstraction and consumes much fewer resources than typical containers.
A state-of-the-art application is \textit{Docker+Wasm}~\cite{dockerwasm}, a special build that makes it possible to run \wasm containers with Docker~\cite{docker2020docker} using the WasmEdge runtime~\cite{wasmedge}.
\wasm has also been used in other server-side applications, including microcontrollers~\cite{gurdeep2019warduino,zandberg2021femto}, trusted execution environments (TEEs)~\cite{menetrey2021twine} and smart contracts~\cite{zheng2020vm,wang2020wana,chen2022wasai}.



With the increase of server-side \wasm applications, studying performance issues of \wasm on the server side becomes highly essential. 
On the one hand, performance degradation (\textit{e.g.}, latency) in server-side applications usually has a more significant impact than in Web applications.
A short latency may not be easily perceived by users in Web applications. But, in some performance-sensitive server applications, it may lead to a decrease in service throughput and cause unexpected economic losses.
Our motivating experiment shows that the latency of \wasm runtimes can significantly affect the throughput of some services (the details will be elaborated in Section~\ref{sec:bg}).
On the other hand, server-side \wasm applications typically run in a standalone runtime system (\textit{e.g.}, WasmEdge~\cite{wasmedge}). 
Unlike major browsers (\textit{e.g.}, Chrome, Safari and Firefox) that have been developed for decades and have powerful optimization mechanisms, existing standalone \wasm runtimes are still in the early development stage. 
Therefore, performance issues of \wasm runtimes are more likely to occur on the server side than on the Web.

However, there is still a lack of research in this area. Existing studies on \wasm performance mainly conducted on the Web environment~\cite{jangda2019not,jangda2019mind,wang2021empowering,yan2021understanding,de2022webassembly,de2021runtime}, while the attention to the server-side \wasm performance is still limited~\cite{spies2021evaluation}.
Moreover, existing research only focuses on the systematic performance gaps between \wasm and native code or JavaScript but lacks attention to \textit{performance issues} in \wasm runtimes.
In particular, performance issues refer to the \textit{abnormal latency} occurring in the \wasm runtimes when running specific applications.
Such performance issues can usually reveal some inappropriate mechanisms (\textit{e.g.}, code optimization, code execution strategy) of specific \wasm runtimes.
Finding performance issues in \wasm runtimes will significantly facilitate the improvement of runtime implementation.


To this end, this paper aims to reveal performance issues in existing standalone \wasm runtimes.
However, there are two main challenges to this task.
First, there are currently a lot of standalone \wasm runtime implementations (more than 30 \wasm runtimes are held on Github~\cite{awesome}). It is hard to analyze each runtime individually. 
The second challenge is determining the \textit{oracle} of performance issues, \textit{i.e.}, there is exactly a performance issue in a \wasm runtime.
Unlike semantic issues causing failure execution or wrong outputs, there is no ground truth of the performance indicator (\textit{i.e.}, execution time of test cases). Furthermore, a longer execution time does not directly indicate a performance issue because this may be caused by the features of the test case instead of the runtime implementation.

To address the first challenge, we adopt the idea of differential testing~\cite{mckeeman1998differential,carlson1979toward,evans2007differential,groce2007randomized}, a typical software testing technique for detecting bugs in a series of comparable systems. The idea is to observe the inconsistency in the outputs of these comparable systems given the same input, which is suitable for testing multiple \wasm runtimes.
However, traditional differential testing approaches only target semantic bugs, which cannot be directly applied to performance issues.
It is infeasible to identify performance issues simply based on the inconsistency in execution time of the same test case since there are systematic performance gaps among different \wasm runtimes.
Therefore, to address the second challenge, we propose a novel differential testing approach \wasmdiff (\textit{\textbf{Wa}sm \textbf{R}untime \textbf{P}erformance \textbf{Diff}erential Testing}) for identifying performance issues in \wasm runtimes. The idea is that in normal cases, the execution time of the same test case on different \wasm runtimes should follow a stable ratio, which we call \ora.
The \ora reflects the systematic performance gaps among different \wasm runtimes.
Thus, for each test case, we first observe the execution time ratio on different \wasm runtimes, then identify an abnormal case in which this ratio significantly deviates from the \ora. For the abnormal case, we further locate which runtime causes the deviation to identify the performance issue.

To evaluate the effectiveness of \wasmdiff, we apply it to identify performance issues in five popular standalone \wasm runtimes (\textit{i.e.}, Wasmer~\cite{wasmer}, Wasmtime~\cite{wasmtime}, Wasm3~\cite{wasm3}, WasmEdge~\cite{wasmedge}, and WAMR~\cite{wamr}) with different settings.
We collect 123 C/C++ programs from the LLVM test suite~\cite{llvmtest} as our test cases. We compile the test programs to \wasm code by Emscripten~\cite{zakai2011emscripten}, then run the \wasm code under each runtime setting and collect their execution time. Based on these data, we identify performance issues in these runtimes by our differential testing approach.
We report the top 10 abnormal cases and summarize seven performance issues in four runtimes.
We further conduct a comprehensive case analysis of these performance issues to reveal their causes. We report these issues to the developers of the corresponding \wasm runtimes, all of which have been confirmed. 
Our code and experiment results are all available\footnote{\url{https://figshare.com/s/f75ddc64d98669ea3abb}}.

The main contributions of this paper are as follows:
\begin{itemize}
    \item We identify the significance of performance issues in server-side \wasm applications, and we conduct the first study on revealing performance issues in server-side \wasm runtimes.
    \item We propose a novel and effective differential testing approach \wasmdiff for identifying performance issues in server-side \wasm runtimes, and we apply it on real-world \wasm runtimes.
    \item We reveal seven unknown performance issues in four \wasm runtimes and further explain their causes with comprehensive case analysis. All the issues have been confirmed by the developers.
\end{itemize}

The rest of this paper is organized as follows. 
Section~\ref{sec:bg} introduces the background of server-side \wasm and illustrates the motivation of our work. Section~\ref{sec:app} describes the design and implementation of \wasmdiff.
Section~\ref{sec:eval} presents our evaluation of \wasmdiff and analysis of identified performance issues.
In Section~\ref{sec:dis}, we discuss threats to validity and future work.
We describe the related work in Section~\ref{sec:re} and finally conclude this work in Section~\ref{sec:con}.

\section{Background and Motivation}\label{sec:bg}

\subsection{\wasm on the Server Side}

\wasm is a low-level bytecode language originally intended for client-side execution in the Web~\cite{haas2017bringing,reiser2017accelerate,wagner2017webassembly}. It serves as the compilation target for applications written in other programming languages such as C/C++, Rust, and Go.
\wasm gains popularity in the Web since it is memory-safe, cross-platform, and provides near-native performance~\cite{hilbig2021empirical}.
Such attributes also make \wasm to be increasingly used on the server side.
In particular, \wasm is a promising solution for running server-side applications in cloud environments~\cite{shillaker2020faasm,gackstatter2022pushing,eriksson2021containerizing,jain2022extending,long2020lightweight}. 
Compared with the traditional Linux containers, \wasm runtimes are safer since they have fewer attack surfaces. \wasm applications are portable across operating systems and CPU architectures. They can also achieve near-native performance by AOT (ahead-of-time) compilation. Furthermore, \wasm consumes much less memory and fewer resources than Linux containers.
In late 2022, Docker~\cite{docker2020docker} announced its support for \wasm with WasmEdge runtime~\cite{wasmedge} called \textit{Docker+Wasm}~\cite{dockerwasm}. This news means that the application of \wasm on the server side has come into practice.

\begin{figure}[t]
  \centering
  \includegraphics[width=0.95\linewidth]{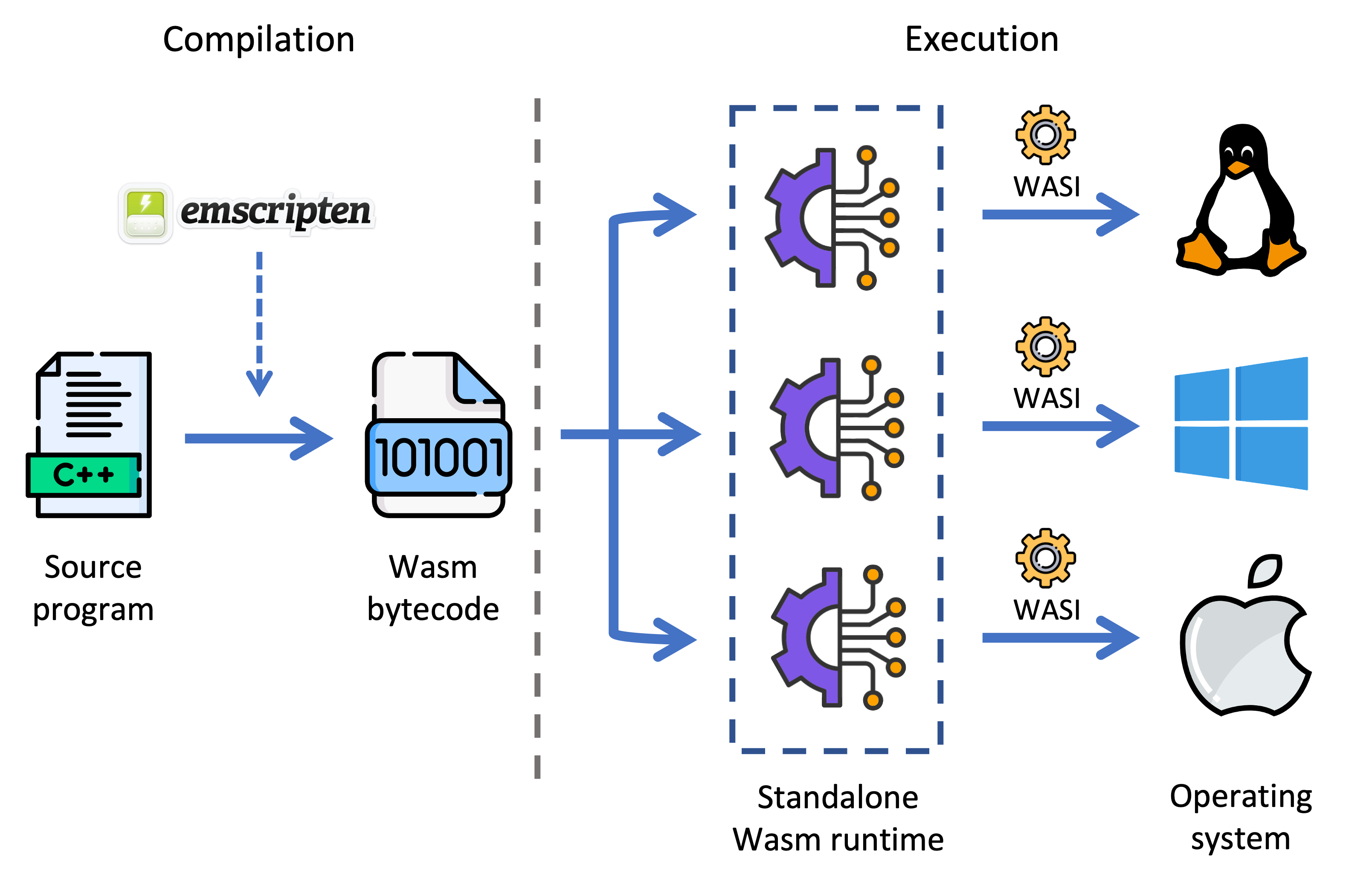}
  \caption{\label{fig:workflow} Typical workflow of a server-side \wasm application.}
\end{figure}

The operating mechanism of \wasm on the server side is different from that in browsers. 
To deploy \wasm applications, we first need to compile the source programs written in high-level languages to \wasm bytecode by specific compilers. For example, Emscripten~\cite{zakai2011emscripten} is a popular compiler that compiles C/C++ to \wasm.
For Web applications, Emscripten generates \wasm module and JavaScript glue code. During the execution, the JavaScript glue code would call into the browser engine (\textit{e.g.}, V8 in Chrome), which would then talk to the operating system.
However, \wasm applications outside browsers need a new way to communicate with the operating system, the WebAssembly System Interface (WASI)~\cite{clark2019standardizing}.
Without a browser engine as runtime, server-side \wasm applications need to run in a standalone runtime system with WASI support. The standalone \wasm runtime works as a sandbox on the host machine, making \wasm applications portable across different platforms.
Figure~\ref{fig:workflow} shows the typical workflow of a server-side \wasm application.

With the increase of server-side \wasm applications, many standalone \wasm runtimes have been developed. Currently, more than 30 standalone \wasm runtimes are held on GitHub~\cite{awesome}.
One representative runtime is Wasmer~\cite{wasmer}, which offers exceedingly lightweight containers executable from cloud, desktop, or IOT devices.
WasmEdge~\cite{wasmedge} is designed by CNCF~\cite{cncf} and integrated with Docker.
Wasmtime~\cite{wasmtime} and WAMR~\cite{wamr} are two other popular runtimes proposed by Bytecode Alliance~\cite{bytecodealliance}. 
The above runtimes all support AOT compilation. There are also runtimes that execute \wasm code in interpreter mode, such as Wasm3~\cite{wasm3}.

However, existing standalone \wasm runtimes are still in the early development stage.
Unlike major browsers (\textit{e.g.}, Chrome, Safari and Firefox) developed for several decades, standalone \wasm runtimes are far from mature and more likely to contain issues, especially performance issues. Performance issues are usually harder to reveal during the testing period than semantic issues, but they can have serious adverse impacts on the application.

\subsection{Impact of Performance Issues}

High performance is a crucial design criterion of \wasm, and it is one of the attributes that make \wasm popular on both the client side and the server side.
However, sometimes there may be performance issues (\textit{i.e.}, abnormal latency) occurring in \wasm runtimes, which is harmful to the reliability of the system.
The impact of performance issues in \wasm runtimes on the server side is even more significant than that on the Web.
Web applications may not be sensitive to a short runtime latency since they usually have client-side I/O much slower than the runtime latency.
On the contrary, server-side applications are usually more performance-sensitive. For example, in some server-side applications with high throughput requirements (\textit{e.g.}, network service), runtime latency may affect the throughput of the application and causes unexpected economic losses.

To study the impact of performance issues in server-side \wasm applications, we conduct a motivating experiment to measure the correlation between \wasm runtime latency and service throughput.
Specifically, we select a real-world \wasm microservice~\cite{microservice} with MySQL database backend as our target application. This microservice is a representative server-side \wasm application supported by \textit{Docker+Wasm}, with WasmEdge as the standalone runtime.
To simulate the performance degradation in \wasm runtime, we insert a loop of numerical computing into the request-handling function of the target service. When receiving a request, the service will execute this loop before handling the request.
In this way, we can introduce runtime latency without changing the functionality of the target service. We can also control the latency time by changing the number of iterations in the loop.
During the experiment, we continuously send requests to the target service from the client machine, and we measure the throughput of the service by \textit{ab}~\cite{ab}, a standard HTTP server benchmarking tool.
We fully occupy the CPU during the request handling to ensure the accuracy of the measured throughput.

\begin{figure}[t]
  \centering
  \subfigure[10,000 requests]{\includegraphics[width=0.49\linewidth]{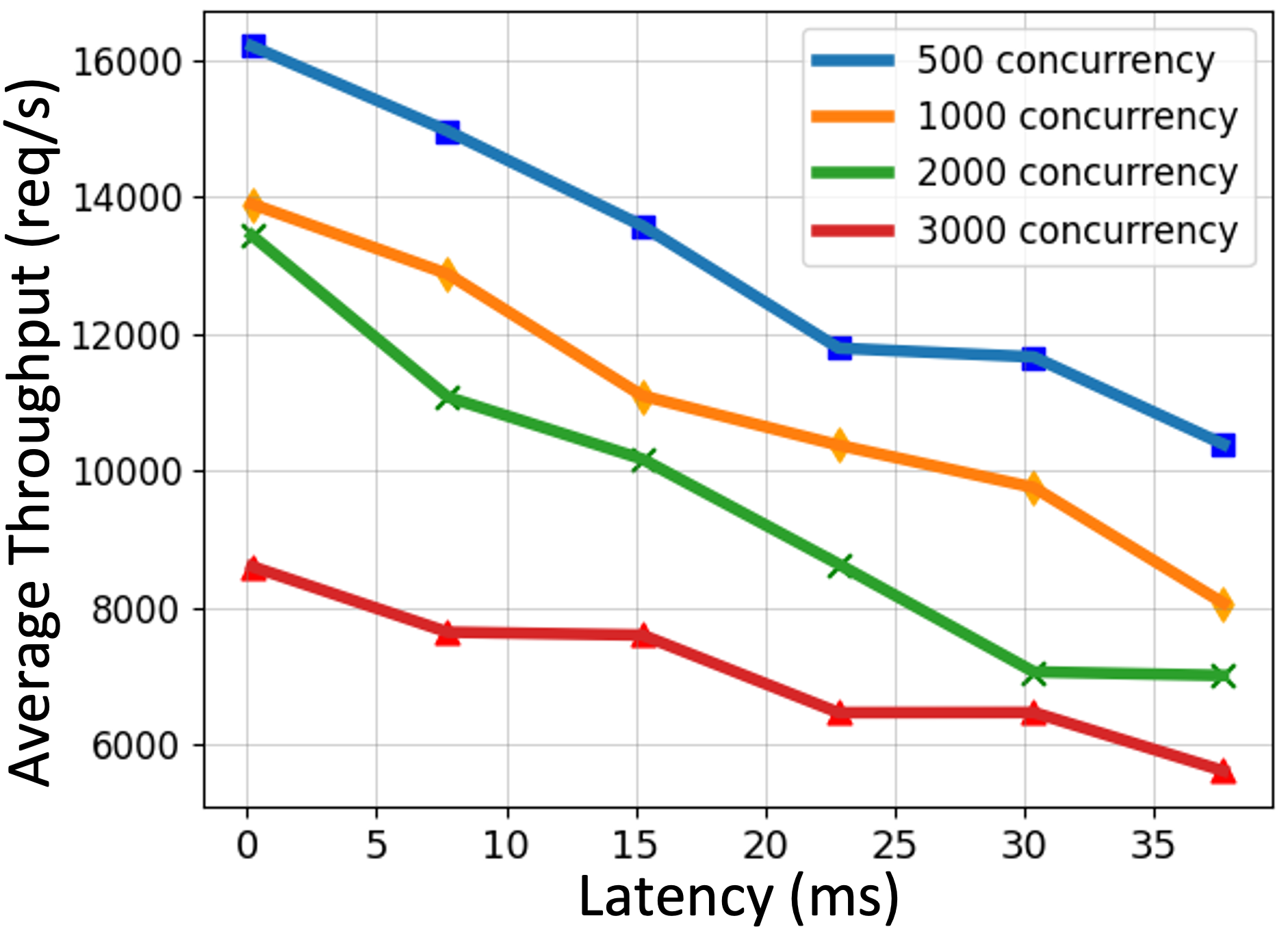}}
  \subfigure[50,000 requests]{\includegraphics[width=0.49\linewidth]{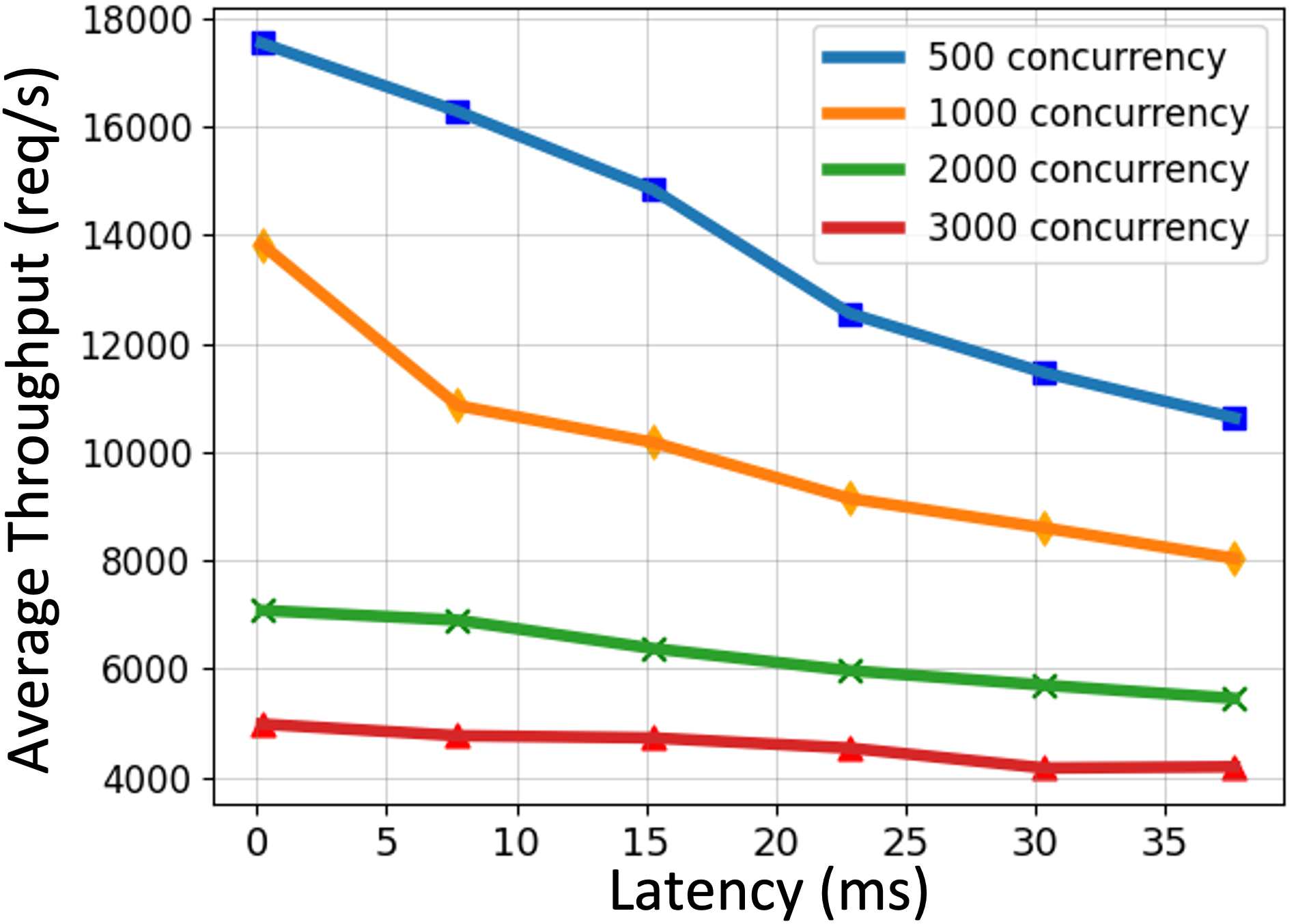}}
  \caption{The impact of WasmEdge runtime latency on service throughput under different concurrency and the total number of requests.}
  \label{fig:throughput}
\end{figure}

Figure~\ref{fig:throughput} shows the correlation between the runtime latency of WasmEdge and service throughput under different concurrency and the total number of requests.
To eliminate the random error of the measurement, we perform seven replicate experiments for each setting and show the average results.
We can find that the runtime latency will cause a severe drop in service throughput under different settings.
Specifically, a short latency of 30ms will result in a 20\% to 50\% drop in service throughput, which is disastrous for high-throughput demanding applications.

Although performance issues can significantly affect the reliability of server-side \wasm applications, there is still a lack of research on performance issues in server-side \wasm runtimes.
Existing studies only focus on the systematic performance gaps between \wasm and native code or JavaScript~\cite{jangda2019not,jangda2019mind,wang2021empowering,yan2021understanding,de2022webassembly,de2021runtime,spies2021evaluation}. To the best of knowledge, none of them has studied performance issues.
Therefore, in this work, we aim to reveal performance issues in existing server-side standalone \wasm runtimes and thus facilitate the improvement of \wasm runtime implementation.

\section{Approach}\label{sec:app}

\begin{figure*}[t]
  \centering
  \includegraphics[width=0.9\linewidth]{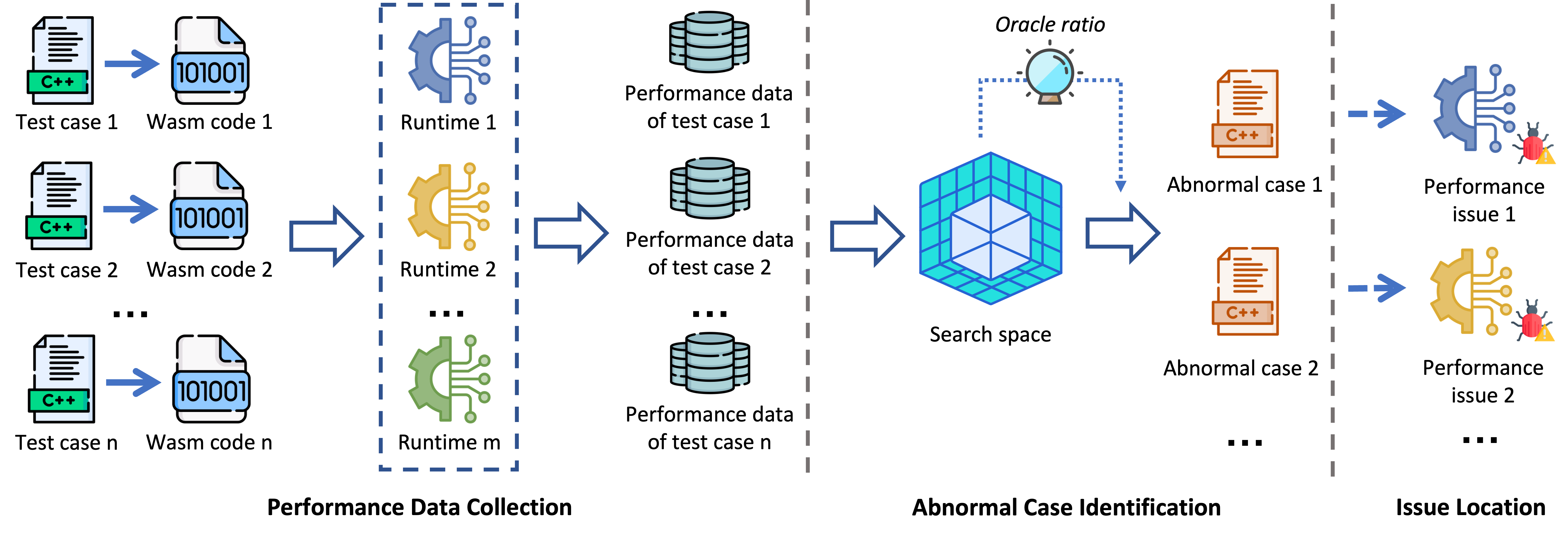}
  \caption{\label{fig:overview} Overall framework of \wasmdiff for identifying performance issues in different standalone \wasm runtimes.}
\end{figure*}

\subsection{Overview}
Finding performance issues in standalone \wasm runtimes is a challenging task. Specifically, there are two main challenges.
First, as we have mentioned above, there are many different implementations of standalone \wasm runtimes. It is time-consuming and labor-intensive to analyze each runtime separately.
Second, it is hard to determine the \textit{oracle} of performance issues, \textit{i.e.}, how to indicate the occurrence of a performance issue. Unlike semantic issues that usually have a ground truth, performance issues cannot be identified by clear criteria. We cannot identify a performance issue simply by observing the execution time of a test case on a \wasm runtime, because the execution time will be affected by the features of the test case.

Therefore, to address the two challenges, we design a novel and effective approach \wasmdiff to identify performance issues in different standalone \wasm runtimes.
We introduce the idea of differential testing~\cite{mckeeman1998differential,carlson1979toward,evans2007differential,groce2007randomized} to solve the first challenge.
Differential testing is a widely-used software testing technique for detecting bugs in multiple comparable systems, by providing the same input to these systems and observing the inconsistency in their execution. It is a suitable solution for our task of testing multiple \wasm runtimes.
However, existing differential testing approaches only target at semantic bugs, which cannot be directly applied to identify performance issues. The challenge of determining the oracle of performance issues still needs to be resolved.
To address this challenge, we introduce a new oracle in \wasmdiff, which is effective in identifying performance issues in different \wasm runtimes.
The key insight is that, in normal cases, the execution time of the same test case on different \wasm runtimes will follow a stable ratio, which we call \ora. Although the execution time of a test case on different \wasm runtimes will be affected by the features of this case and the systematic performance gaps among different runtimes, the \ora can always be an indicator of normal performance.
Therefore, we can identify an abnormal case in which the execution time ratio on different runtimes significantly deviates from the \ora. The abnormal cases can indicate performance issues in some \wasm runtimes, and we further locate the specific runtime in which the performance issues occurred.

Figure~\ref{fig:overview} illustrates the overall framework of \wasmdiff for identifying performance issues in different standalone \wasm runtimes.
Specifically, our approach can be divided into three phases: 
(1) \textit{Performance data collection}. We execute each test case on multiple \wasm runtimes and collect the performance data; 
(2) \textit{Abnormal case identification}. We determine the \ora based on the performance data of each test case and identify abnormal cases; 
(3) \textit{Performance issue location}. We analyze the performance data of the abnormal cases to locate the \wasm runtime with performance issues.
In the following subsections, we will elaborate on the details of our design and implementation of these three phases.

\subsection{Performance Data Collection}
In order to identify performance issues in different standalone \wasm runtimes, we first need to collect performance data of various test cases on these runtimes. This phase can be further divided into three sub-steps: test case selection, \wasm code execution, and performance data recording.

For test case selection, we need to consider the types of source programs that can be well supported by standalone \wasm runtimes. Currently, \wasm has relatively complete support for source programs written in C/C++ and Rust~\cite{hilbig2021empirical}. Therefore, it is appropriate to select C/C++ or Rust programs as test cases. Furthermore, we should select test cases that are more likely to trigger performance issues in \wasm runtimes, \textit{e.g.}, source programs from some benchmark suites for performance testing.

For each test case, we compile the source program to \wasm code and then execute it on different standalone \wasm runtimes. 
In this step, we need to ensure the correctness of the execution results of the test cases. We exclude test cases where the execution result is incorrect or a runtime error occurs during execution, because it is meaningless to evaluate the performance of such cases.
In order to eliminate random errors of code execution, we execute each test case multiple times on each runtime and record the average value of performance data. 
The number of executions can be customized according to the requirements for test efficiency.

During the execution of a test case, we need to record the performance data of this test case on each \wasm runtime for differential testing. In this step, we need to consider what performance data to record.
The most intuitive idea is that for each test case, record the time of the whole process of its \wasm code running on each runtime. 
But in order to better locate and analyze the identified performance issues, we record the performance data with finer granularity.
Specifically, the whole running process of \wasm code on a runtime consists of three stages: \textit{runtime initialization}, \textit{\wasm code loading}, and \textit{code execution}. 
Runtime initialization is where the \wasm runtime starts and prepares the code execution environment. Then in the \wasm code loading stage, the runtimes in AOT mode will first compile the \wasm code to executable binary, while the runtimes in interpreter mode just load the \wasm code into memory. Finally, the runtime performs code execution.
Therefore, for each test case, we record the time of these three stages when it runs on each \wasm runtime. For implementation, we use the Linux \textit{perf} tool~\cite{perf} to find the start and end positions of these three stages and record the time stamps during the test case run.
We also record the total time of the whole running process.

\subsection{Abnormal Case Identification}
In this phase, we aim to identify abnormal test cases based on the performance data we have collected. 
For the convenience of data analysis, we only take the \textit{total time} as the performance indicator in this phase (For consistency, we refer to “total time” as “execution time” in the following). The time of the three running stages will be used for further analysis of the identified performance issues.

As we have mentioned above, the key idea of identifying abnormal cases is to observe the execution time ratio on different \wasm runtimes for each case, and the cases where this ratio significantly deviates from the \ora are considered abnormal cases.
To this end, we need to solve two problems: (1) represent the execution time ratio for each test case; (2) determine the \ora.

For the first problem, an appropriate solution is test case vectorization.
Specifically, for each test case, we create a time vector to represent the execution time ratio according to the execution time of this case on each \wasm runtime. 
For example, if the execution time of case $x$ on 3 \wasm runtimes is 1s, 2s, and 3s respectively, the time vector of case $x$ can be represented as $[1,2,3]$. 
However, the time vectors of different test cases cannot be directly compared since the execution time is related to the features of the case itself.
Therefore, to make the time vectors of different test cases comparable, we need to normalize the time vectors for all test cases. In this way, the difference in execution time caused by different test cases can be eliminated. Test cases with the same execution time ratio will have the same normalized time vectors. For example, the time vector $[2,4,6]$ of case $y$ will be the same as that of case $x$ after normalization.

For the second problem, the ideal solution is that we already know the \ora. Unfortunately, the \ora cannot be predicted in advance, since the normal performance of each \wasm runtime is currently unknown.
The current optimal solution to this problem is to estimate the \ora according to the execution time ratio of the existing test cases.
Specifically, we have mapped the execution time ratio of all the test cases to the same search space by test case vectorization and normalization.
We treat the center (\textit{i.e.}, mean vector) of all the normalized time vectors as the vector of the estimated \ora.
Assuming that most test cases are normal cases where the execution time ratio is similar to the \ora, when there are enough test cases, our estimated \ora will approach the ideal \ora.

Thus, for each test case, we can calculate the distance between its normalized time vector and the vector of the estimated \ora in the search space.
Although the estimated \ora will be affected by the abnormal cases, in general, a greater distance means a higher anomaly degree for a test case.
Therefore, we rank all the test cases according to this distance, and we identify abnormal cases from the top of this ranking.

\subsection{Performance Issue Location}
After we find an abnormal case, we need to locate which \wasm runtime caused this anomaly, \textit{i.e.}, in which runtime a performance issue occurred.
To this end, we analyze the impact of each \wasm runtime on this anomaly respectively, based on the execution time of this abnormal case on each runtime.
According to our strategy for identifying abnormal cases, the time vector of an abnormal case is relatively far from the vector of the estimated \ora. This distance is mainly caused by the abnormal execution time of this case on some \wasm runtimes, \textit{i.e.}, some dimensions with an abnormal value in the time vector.
Therefore, we can evaluate the effect of the value of each dimension on this distance separately.
Specifically, we adjust the value of one dimension to make the time vector closest to the vector of the estimated \ora. We repeat this operation for each dimension and record the value that needs to be adjusted, which we call \dd.
This \dd reflects the impact of the corresponding \wasm runtime on this abnormal case. The larger \dd means that this anomaly is more likely to be caused by this runtime. Thus, we treat the \wasm runtime with the largest \dd as the issue-related runtime.

Hence, for each abnormal case, we can locate the \wasm runtime in which the performance issue occurred by the above solution.
It is worth noting that \wasmdiff is only a heuristic approach, and there may be other solutions for this problem. We just propose a feasible solution and hope our work can inspire more refined approaches in the future.
We will show the effectiveness of \wasmdiff in the next section.

\section{Evaluation and Analysis}\label{sec:eval}
To evaluate the effectiveness of \wasmdiff, we apply it on several real-world standalone \wasm runtimes.
In this section, we aim to answer the following research questions:
\begin{itemize}
    \item \textbf{RQ1:} How does \wasmdiff perform in identifying performance issues in real-world standalone \wasm runtimes?
    \item \textbf{RQ2:} What are the causes of the identified performance issues, and how can we verify them?
    \item \textbf{RQ3:} What is the computational overhead of differential testing in \wasmdiff?
\end{itemize}

\subsection{Experiment Settings}

\textbf{Test Case Selection.}
As described in Section~\ref{sec:app}, it is appropriate to select test cases written in source languages that are well supported by \wasm and that are more likely to trigger performance issues.
Therefore, we select 141 C/C++ programs with a total of over 30,000 lines of code from the LLVM test suite~\cite{llvmtest}, which contains various benchmark programs for evaluating LLVM compilation performance.
We select the test cases from the \textit{SingleSource/Benchmarks/} directory of the test suite, since the programs in this directory can be directly compiled to \wasm code without modification.
Table~\ref{tab:testcases} shows the information of our selected test cases, consisting of 14 benchmarks. One of the benchmarks is Polybench~\cite{polybench}, which is a widely-used benchmark suite for \wasm performance evaluation in previous studies~\cite{jangda2019not, jangda2019mind, spies2021evaluation}.
We compile the source programs to \wasm code by Emscripten (version 3.1.24) with the optimization level of \textit{O2}.
We exclude those test cases that cannot be compiled successfully or a runtime error occurs during execution. Finally, we collect the results on the remaining 123 test cases.

\begin{table}[t!]
\centering
\caption{Information of our test cases from the LLVM test suite.}
\begin{threeparttable}
\resizebox{\linewidth}{!}{
\begin{tabular}{lcc|lcc}
 \toprule
 \textbf{Benchmark} & \textbf{\#Program} & \textbf{\#LOC\tnote{*}} & \textbf{Benchmark} &  \textbf{\#Program} & \textbf{\#LOC\tnote{*}}\\
 \midrule
 Adobe-C++ & 6 & 1,615 & Misc-C++ & 7 & 1,322 \\
 BenchmarkGame & 8 & 486 & Misc-C++-EH & 1 & 16,817 \\
 CoyoteBench & 4 & 1,471 & Polybench & 30 & 4,364 \\
 Dhrystone & 2 & 642 & Shootout & 14 & 573 \\
 Linpack & 1 & 693 & Shootout-C++ & 25 & 783 \\
 McGill & 4 & 956 & SmallPT & 1 & 96 \\
 Misc & 27 & 5,052 & Stanford & 11 & 1,135 \\
 \midrule
  &  &  & \textbf{Total} & 141 & 36,005 \\
 \bottomrule
\end{tabular}}
\begin{tablenotes}
    \item[*] LOC: lines of code.
\end{tablenotes}
\end{threeparttable}
\label{tab:testcases}
\end{table}

\textbf{\wasm Runtime Selection.}
Although there are many server-side standalone \wasm runtimes, it is better to select some representative \wasm runtimes as test targets.
Since most standalone \wasm runtimes are open source on GitHub, we select target runtimes based on their \textit{popularity} and \textit{activity} on Github.
For popularity, we select runtimes with the top number of Github stars. For activity, we exclude those unmaintained runtimes, \textit{i.e.}, the last commit was more than one year ago.
Finally, we select five representative standalone \wasm runtimes: Wasmer~\cite{wasmer}, Wasmtime~\cite{wasmtime}, Wasm3~\cite{wasm3}, WasmEdge~\cite{wasmedge}, and WebAssembly Micro Runtime (WAMR)~\cite{wamr}.
Table~\ref{tab:runtimes} shows the information of these \wasm runtimes. We select the latest version of each runtime for testing.
For Wasmer, Wasmtime, and WasmEdge, we test them under AOT mode. Although WasmEdge also provides interpreter mode, the performance is extremely poor, so we only test WasmEdge with AOT mode. For Wasm3, we test it on the default interpreter mode and another setting with \textit{--compile} option, where the lazy optimization of \wasm code will be disabled. For WAMR, we test it on both the interpreter mode and AOT mode. Hence, we finally have seven runtime settings.

\begin{table}[t!]
\centering
\caption{Information of \wasm runtimes for testing.}
\begin{threeparttable}
\begin{tabular}{lccc}
 \toprule
 \textbf{Runtime} & \textbf{\#GitHub Stars\tnote{*}}  & \textbf{Test Version} & \textbf{Execution Mode} \\
 \midrule
 Wasmer & 15.1k & 3.2.0 & AOT \\
 Wasmtime & 12.1k & cli 8.0.0 & AOT \\
 Wasm3 & 6k & v0.5.0 & Interpreter \\
 WasmEdge & 5.9k & 0.12.0 & AOT \\
 WAMR & 3.7k & 1.1.2 & Interpreter/AOT \\
 \bottomrule
\end{tabular}
\begin{tablenotes}
    \item[*] Statistics of Github stars is by April 2023.
\end{tablenotes}
\end{threeparttable}
\label{tab:runtimes}
\end{table}

\textbf{Experiment Environment.}
All our experiments are running on a server with an Intel(R) Core(TM) i5-9500T 2.20GHz CPU and 32GB DDR4 memory. The operating system of the server is 64-bit Ubuntu 20.04.1 SMP with Linux kernel version of 5.15.0-56-generic.

\subsection{RQ1: Results of Identifying Performance Issues}
We run each test case 10 times under each runtime setting and collect the performance data averaged over the 10 runs. Then we apply \wasmdiff on all runtime settings and obtain the results.
According to our approach, we identify abnormal cases and then locate performance issues in specific \wasm runtimes based on their \dd in each case. The larger \dd indicates that the case performance on the corresponding runtime is with the higher anomaly.
Therefore, we rank the identified abnormal cases based on the descending order of the \dd of the issue-related runtime.
Table~\ref{tab:abnormal} shows the results of the top 10 abnormal cases. 
We only report the top 10 abnormal cases since we just rank the cases without setting a specific threshold for abnormal cases. We design this strategy because our goal is to reveal some unknown performance issues in existing standalone \wasm runtimes, instead of finding all the performance issues. Actually, it is impossible to find all the performance issues, because there is currently no ground truth of performance issues that can be verified.

\begin{table*}[t!]
\centering
\caption{\textit{Deviation degree} of each runtime setting on the top 10 abnormal cases.}
\begin{tabular}{lccccccc}
 \toprule
 \textbf{Case} & \textbf{Wasmer}  & \textbf{Wasmtime} & \textbf{Wasm3} & \textbf{Wasm3\_compile} & \textbf{WasmEdge} & \textbf{WAMR} & \textbf{WAMR\_AOT}\\
 \midrule
 \texttt{BenchmarkGame/fasta.c} & \colorbox{lightgray}{0.702} & 0.113 & -0.248 & -0.244 & 0.082 & -0.270 & 0.081 \\
 \texttt{Shootout/methcall.c} & -0.051 & -0.028 & -0.164 & -0.164 & \colorbox{lightgray}{0.502} & 0.044 & -0.014 \\
 \texttt{Shootout-C++/methcall.cpp} &	-0.036 & -0.031 & -0.126 & -0.128 & \colorbox{lightgray}{0.415} & 0.072 & -0.009 \\
 \texttt{Shootout/random.c}	& 0.075 & \colorbox{lightgray}{0.315} & -0.060 & -0.060 & 0.079 & -0.026 & 0.101 \\
 \texttt{Shootout-C++/random.cpp} & 0.096 & \colorbox{lightgray}{0.309} & -0.063 & -0.063 & 0.098 & -0.036 & 0.121 \\
 \texttt{Polybench/2mm.c} & -0.038 & -0.039 & -0.151 & -0.149 & -0.035 & \colorbox{lightgray}{0.268} & 0.003 \\
 \texttt{Polybench/gemm.c} & -0.038 & -0.041 & -0.145 & -0.153 & -0.036 & \colorbox{lightgray}{0.267} & 0.007 \\
 \texttt{Polybench/3mm.c} & -0.037 & -0.040 & -0.145 & -0.140 & -0.034 & \colorbox{lightgray}{0.261} & 0.005 \\
 \texttt{Misc/flops-8.c} & -0.019 & 0.012 & -0.142 & -0.142 & -0.009 & \colorbox{lightgray}{0.251} & 0.015 \\
 \texttt{Misc/flops-4.c} & \colorbox{lightgray}{0.234} & -0.003 &	-0.127 & -0.127	& -0.019 & 0.168 & 0.001 \\ 
 \bottomrule
\end{tabular}
\label{tab:abnormal}
\end{table*}

The values in Table~\ref{tab:abnormal} represent the \dd of each \wasm runtime setting on the top 10 abnormal cases. A positive value means that the execution time of this case on this \wasm runtime is higher than the expected value according to the \ora, while a negative value means that the execution time is lower than the expected value. Since we aim to identify performance issues (\textit{i.e.}, the execution time is abnormally higher than expected), we only focus on the \dd with positive values.
For each abnormal case, the issue-related runtime is marked with a gray background in the table. 
We can observe that among the 10 abnormal cases, four cases are caused by WAMR with interpreter mode, and the other six cases are caused by Wasmer, Wasmtime and WasmEdge (two cases on each runtime). There are also other abnormal cases caused by Wasm3, which are not shown in the table.

The results indicate that performance issues are common in existing popular standalone \wasm runtimes, which need our attention. 
We will further conduct a detailed case analysis to reveal the causes of these performance issues.

\subsection{RQ2: Case Analysis}
In order to verify the identified performance issues and further facilitate the improvement of \wasm runtime implementation, it is critical to analyze the causes of these performance issues.
Unfortunately, since the performance issues we identified are all unknown issues, there are no ground truths that can be directly used for verification.
Therefore, we conduct a manual analysis of these abnormal cases. Specifically, we analyze each case in three steps: \textit{abnormal stage location}, \textit{fine-grained cause location}, and \textit{cause verification}.

In the first step, we locate the running stage where the abnormal latency occurs. As mentioned in Section~\ref{sec:app}, we have collected the time of the three running stages (runtime initialization, \wasm code loading, and code execution) for each test case. Thus, we can locate the abnormal stage based on these performance data. 
Similarly, we apply \wasmdiff on the data of these three stages respectively and identify the abnormal stage where the issue-related runtime (\textit{e.g.}, for case \texttt{fasta.c}, the issue-related runtime is Wasmer) holds the largest \dd.
We find that in all 10 abnormal cases, the abnormal latency occurs in the code execution stage. This means that the performance issues we identified are all caused by the code execution mechanism of the corresponding runtimes.

This finding indicates that we can locate fine-grained causes of the performance issues by analyzing the source code of the abnormal cases.
Therefore, in the second step, we aim to find out which part of the code is executing with an abnormal latency.
To this end, for each abnormal case, we make a series of case reduction, and we rerun the reduced cases on all the \wasm runtimes to observe the changes in the execution time ratio. Specifically, we generate a reduced case by deleting a code snippet (\textit{e.g.}, a statement, a loop, or a branch). If the execution time ratio of the reduced case changes to the normal level (\textit{i.e.}, close to the \ora), it means that the performance issue is likely to be caused by the deleted code snippet, which we call \ics.

To verify the causes of the performance issues, we further create some new test cases that contain the same function as the \ics. We run the new test cases on all the \wasm runtimes and observe whether the performance issues will be reproduced. If the performance issues can be reproduced, it means that the causes we found can be verified.
Finally, we report the performance issues and their causes to the developers of the corresponding \wasm runtimes.

Overall, we summarize 7 performance issues for the 10 abnormal cases, all of which have been confirmed by the developers.
Table~\ref{tab:issue-summary} shows the summary of these performance issues.
Next, we will explain the performance issues of each \wasm runtime separately.

\begin{table*}[t!]
\centering
\caption{Summary of performance issues related to the 10 abnormal cases.}
\begin{tabular}{lcccc}
 \toprule
 \textbf{Case} & \textbf{Related Runtime}  & \textbf{Issue ID} 
 & \textbf{Cause of Performance Issue} & \textbf{Status} \\
 \midrule
 \texttt{BenchmarkGame/fasta.c} & Wasmer & $\#3784$ & Improper implementation of \texttt{fd\_write} & Confirmed \\
 \texttt{Misc/flops-4.c} & Wasmer  & $\#3821$ & Version issue of the \textit{Cranelift} code generator & Confirmed \\ 
 \midrule
 \texttt{Shootout/methcall.c} & WasmEdge & $\#2444$ & Improper handling when invoking function pointer & Confirmed \\
 \texttt{Shootout-C++/methcall.cpp} & WasmEdge & $\#2442$ & Improper handling of virtual function & Confirmed \\
 \midrule
 \texttt{Shootout/random.c}	&  Wasmtime & \multirow{2}*{$\#6287$} & \multirow{2}*{Insufficient optimization for division and modulo} & \multirow{2}*{Confirmed} \\
 \texttt{Shootout-C++/random.cpp} & Wasmtime  &  &  & \\
 \midrule
 \texttt{Polybench/2mm.c} & WAMR & \multirow{3}*{$\#2175$} & \multirow{3}*{Insufficient optimization for matrix multiplications} & \multirow{3}*{Confirmed}\\
 \texttt{Polybench/gemm.c} & WAMR &  &  & \\
 \texttt{Polybench/3mm.c} & WAMR &  &  & \\
 \hline 
 \texttt{Misc/flops-8.c} & WAMR  & $\#2167$ & Insufficient optimization for complex arithmetic expressions & Confirmed \\
 
 \bottomrule
\end{tabular}
\label{tab:issue-summary}
\end{table*}

\begin{figure}[t]
  \centering
  \subfigure[Issue-related code snippet of \texttt{fasta.c}.]{\includegraphics[width=0.98\linewidth]{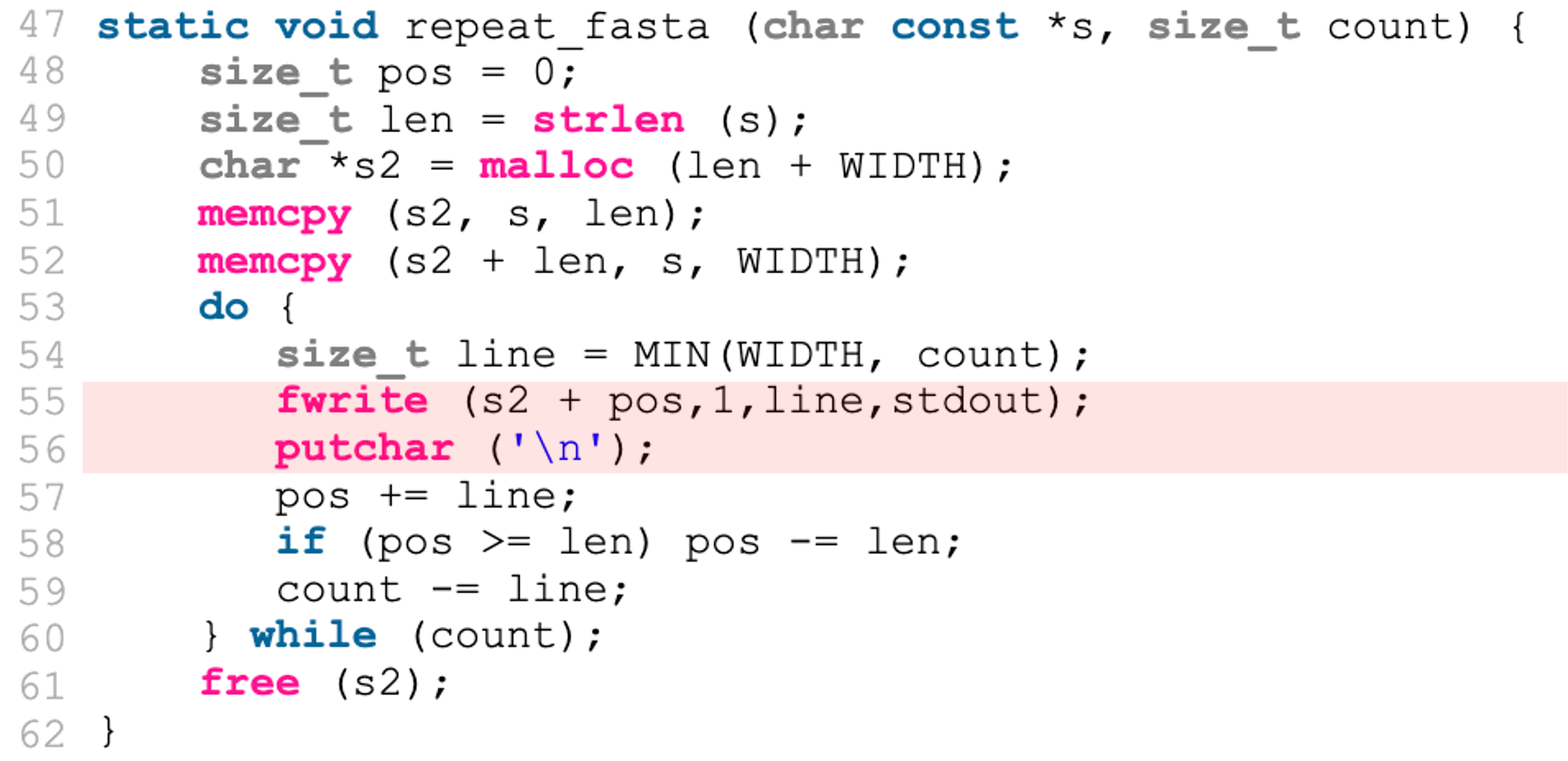}\label{fig:case1-1}}\\
  \subfigure[A new test case that can reproduce Issue $\#3784$.]{\includegraphics[width=0.98\linewidth]{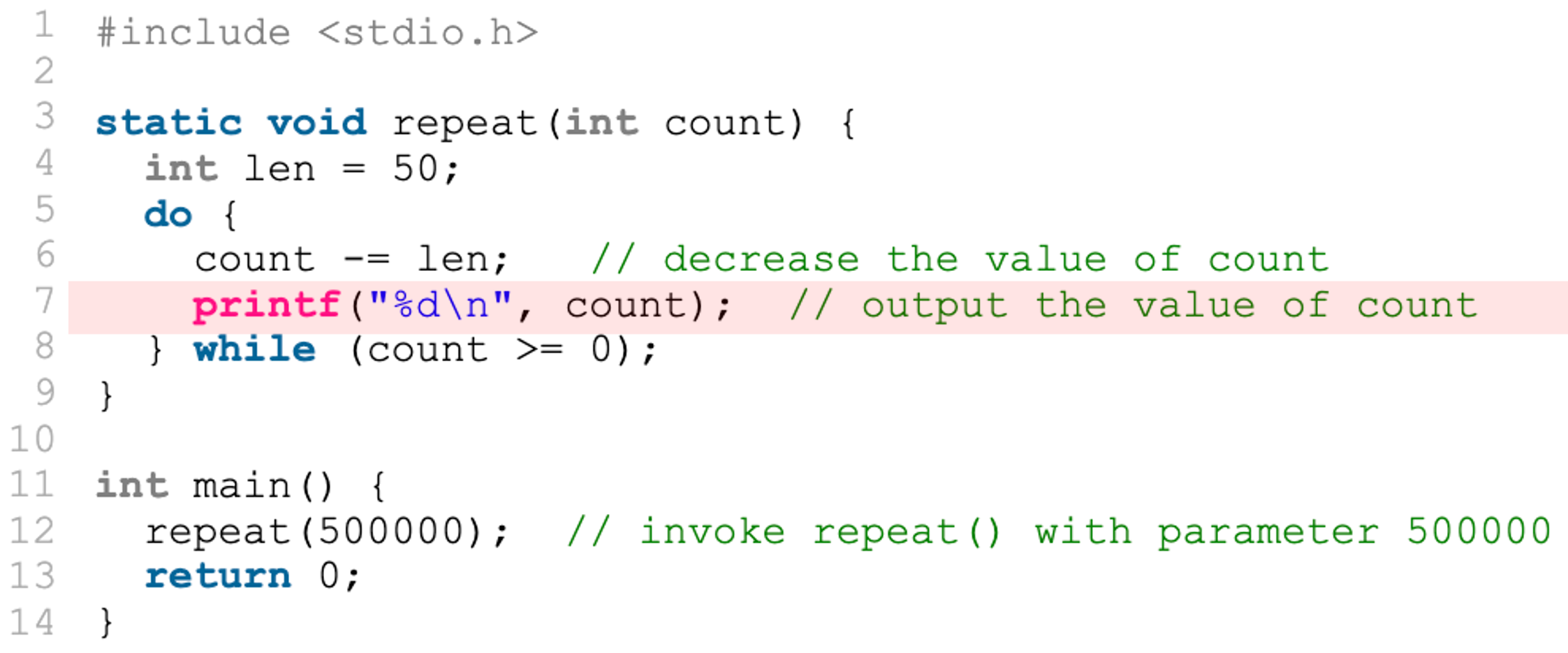}\label{fig:case1-2}}
  \caption{Test cases related to Issue $\#3784$ of Wasmer.}
  \label{fig:case1}
\end{figure}

~\\\textbf{Wasmer.}
In Issue $\#3784$, we find that the core function in the abnormal case \texttt{fasta.c} is \texttt{repeat\_fasta}, as shown in Figure~\ref{fig:case1-1}. This function prints the characters of the string $s$ repeatedly, and it stops when the total number of characters printed is $count$. 
We further locate the \ics, which accounts for the majority of the case execution time at lines 55-56 by case reduction. Here the program invokes two C standard I/O functions \texttt{fwrite} and \texttt{putchar} in a loop. When we delete these two lines of code, the execution time of this case on Wasmer will go back to normal. Therefore, this performance issue is probably caused by improper I/O implementation of Wasmer. 
To verify this cause, we create a new test case that also includes a standard I/O function \texttt{printf} in a loop, as shown in Figure~\ref{fig:case1-2}. We find that the performance issue of Wasmer can be reproduced in this case.
Then, we check the source code of I/O implementation in Wasmer (in \textit{wasmer/lib/wasi/src/syscalls/wasi/fd\_write.rs}), delete the code snippet of setting written size and rebuild Wasmer. We find that the performance issue will not occur after rebuilding.
Therefore, the cause of improper I/O implementation in Wasmer can be verified.

In Issue $\#3821$, the abnormal case \texttt{flops-4.c} is a program that calculates the integral of $cos(x)$ using the trapezoidal method. We find that the \ics of this case is a statement that performs arithmetic operations. Thus, the performance issue may be related to Wasmer's improper handling of such operations. Specifically, \wasm runtimes in AOT mode will generate executable machine code for the input \wasm code before execution. The default code generator of Wasmer is \textit{Cranelift}~\cite{cranelift}. When we change the code generator to the LLVM backend and rerun this case on Wasmer, the performance is back to normal.
However, Wasmtime also uses \textit{Cranelift} as the default code generator but no performance issue occurs, which indicates that the issue is caused by the current version of \textit{Cranelift} in Wasmer.

These two performance issues of Wasmer have been confirmed by the developers and marked as milestones for the development of the next version.

\begin{figure}[t]
  \centering
  \includegraphics[width=\linewidth]{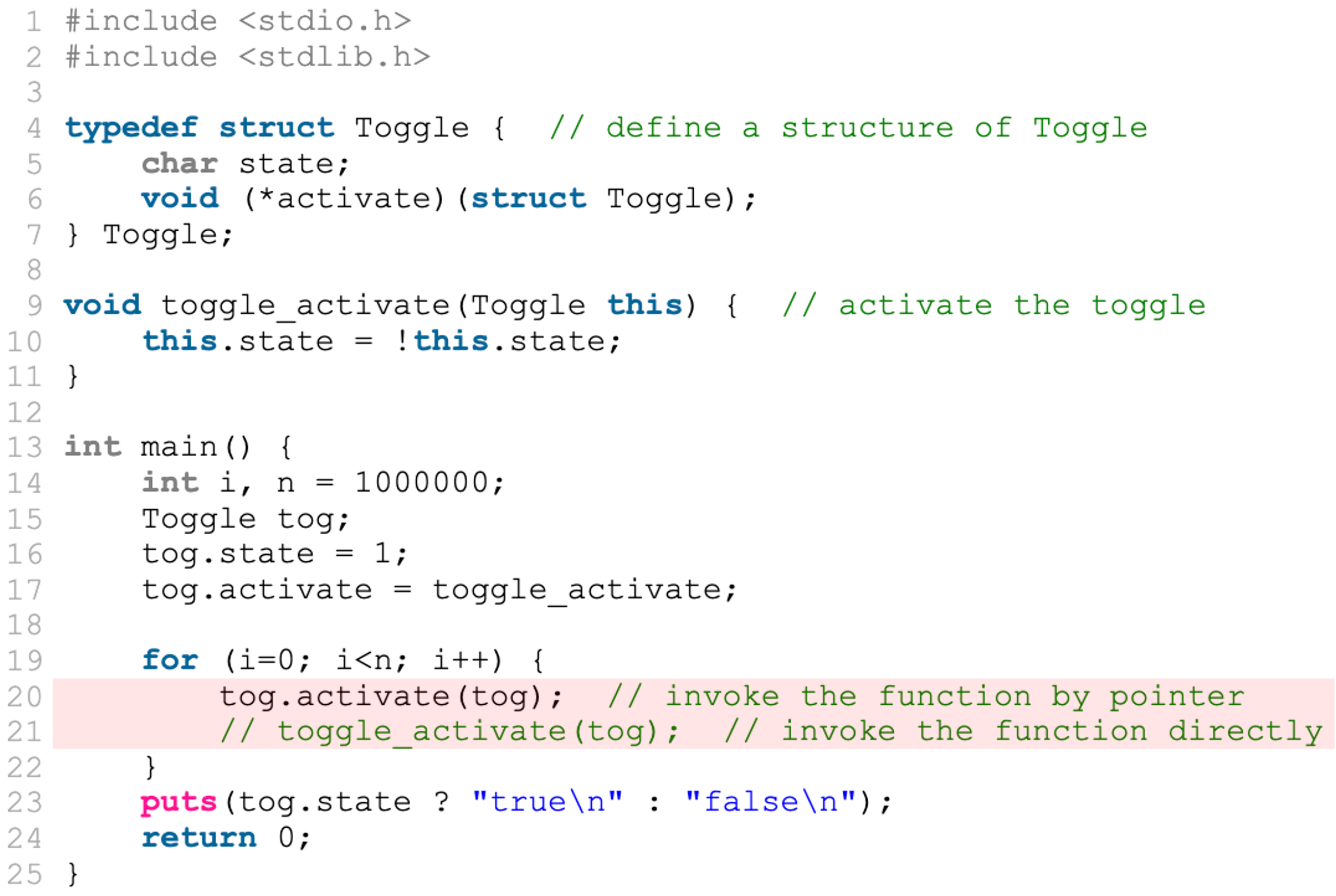}
  \caption{\label{fig:case3} Simplified \texttt{methcall.c} related to Issue $\#2444$ of WasmEdge.}
\end{figure}

~\\\textbf{WasmEdge.}
In Issue $\#2444$, the abnormal case \texttt{methcall.c} defines a structure named $Toggle$, and it invokes a function to activate the toggle repeatedly. For the convenience of explaining the issue, we create a simplified \texttt{methcall.c}, as shown in Figure~\ref{fig:case3}. In this case, we locate the \ics at line 20, where the program invokes the function \texttt{toggle\_activate} via the function pointer \texttt{activate} defined in the structure $Toggle$. 
However, when we remove the code of this line and invoke the function \texttt{toggle\_activate} directly (as shown in line 21), the performance issue of WasmEdge will not show up again.
The results indicate that this performance issue is caused by the improper handling of WasmEdge when invoking a function pointer.

The abnormal case \texttt{methcall.cpp} in Issue $\#2442$ implements the same function as \texttt{methcall.c}, but written in C++.
Due to differences in syntax of C and C++, the function pointer \texttt{activate} in \texttt{methcall.c} is defined as a virtual function reference \texttt{virtual bool\& activate()} in \texttt{methcall.cpp}. The performance issue in this case also occurs when invoking \texttt{activate}.
Therefore, we find that WasmEdge also has improper handling of a virtual function.
These two performance issues are also confirmed by the developers of WasmEdge.

\begin{figure}[t]
  \centering
  \subfigure[Issue-related code snippet of \texttt{random.c}.]{\includegraphics[width=\linewidth]{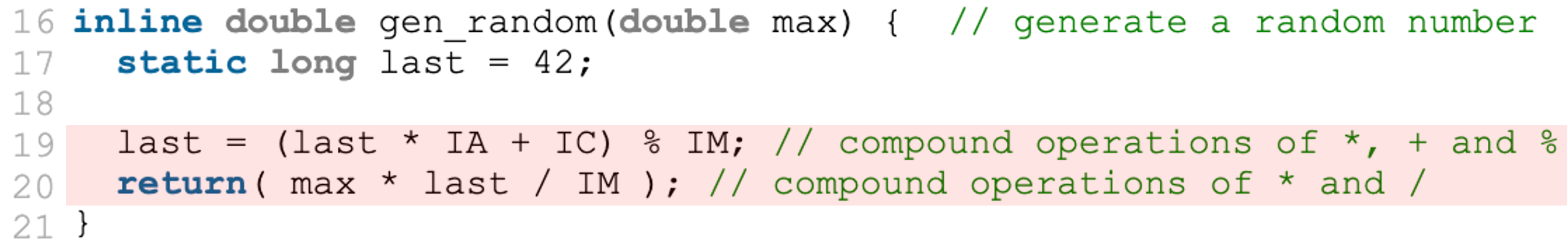}\label{fig:case5-1}}\\
  \subfigure[A new test case that can reproduce Issue $\#6287$.]{\includegraphics[width=\linewidth]{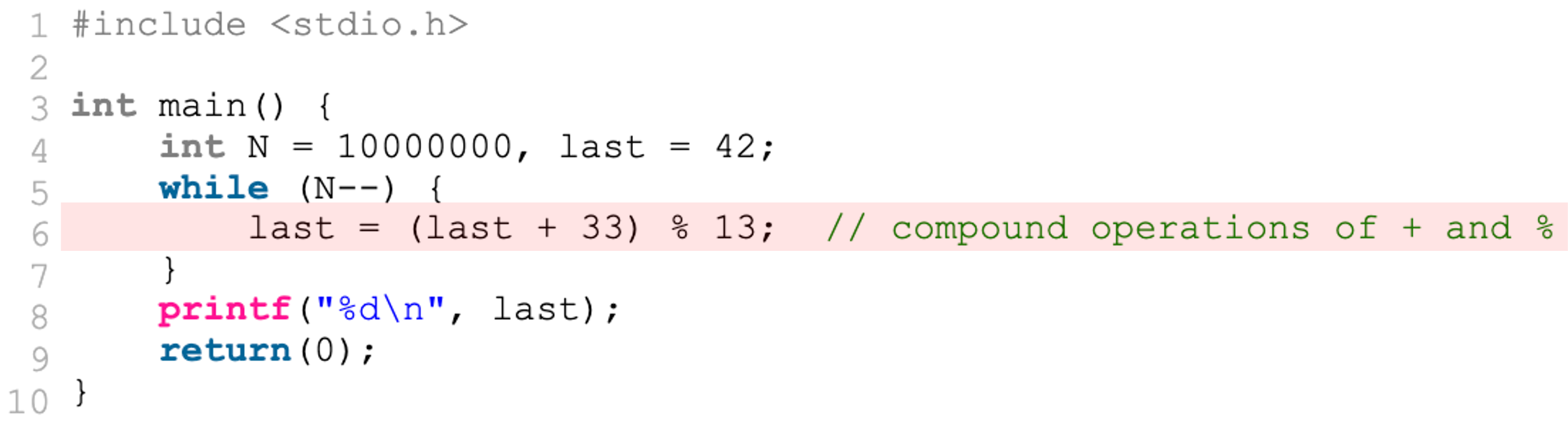}\label{fig:case5-2}}
  \caption{Test cases related to Issue $\#6287$ of Wasmtime.}
  \label{fig:case5}
\end{figure}

~\\\textbf{Wasmtime.}
The abnormal cases \texttt{random.c} and \texttt{random.cpp} reveal the same performance issue $\#6287$ of Wasmtime. The core functions of the two programs are generating a random number by some compound operations, as shown in Figure~\ref{fig:case5-1}. We locate the \ics at lines 19-20, which means that Wasmtime may handle such compound operations improperly. We then create another test case with a similar function, as shown in Figure~\ref{fig:case5-2}, and find that the performance issue will be reproduced. We further create more test cases that contain different compound operations, and we find that this performance issue of Wasmtime only occurs when division and modulo are included.
We report this performance issue to the developers of Wasmtime. They confirm this issue and admit that the optimization of division and modulo is currently not well supported by Wasmtime.

\begin{figure}[t]
  \centering
  \includegraphics[width=\linewidth]{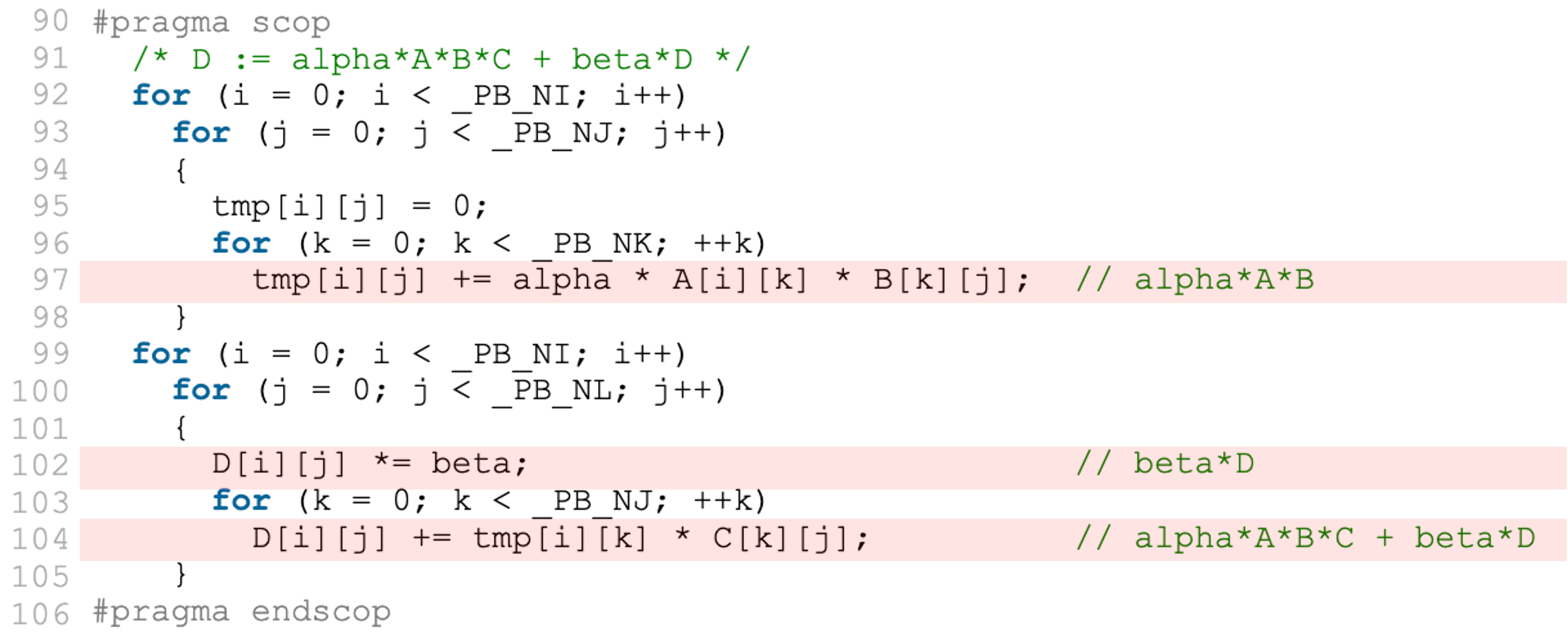}
  \caption{\label{fig:case6} Issue-related code snippet of \texttt{2mm.c} in Issue $\#2175$ of WAMR.}
\end{figure}

\begin{figure}[t]
  \centering
  \includegraphics[width=\linewidth]{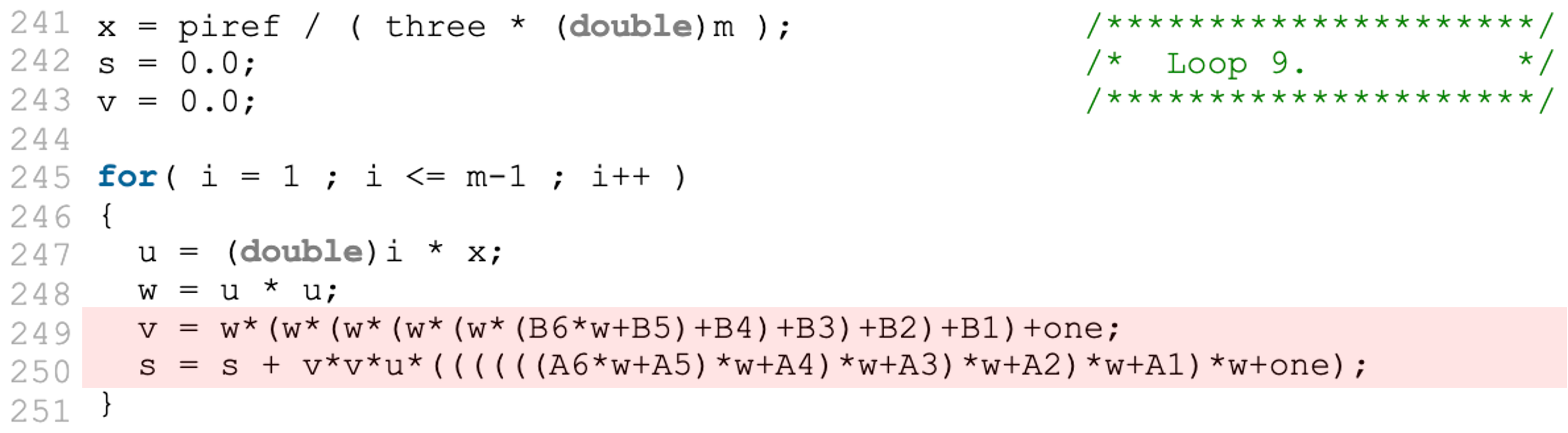}
  \caption{\label{fig:case7} Issue-related code snippet of \texttt{flops-8.c} in Issue $\#2167$ of WAMR.}
\end{figure}

~\\\textbf{WAMR.} 
The three abnormal cases \texttt{2mm.c}, \texttt{gemm.c}, and \texttt{3mm.c} from Polybench reflect the same performance issue $\#2175$ of WAMR in interpreter mode. The functions of these three programs are all matrix multiplication. Figure~\ref{fig:case6} shows the core computations in \texttt{2mm.c}, where the program performs the operation $alpha * A * B * C + beta * D$ on matrices $A$, $B$, $C$, $D$. We locate the \ics at lines 97, 102, and 104, which are the statements of matrix multiplication. We also obtain similar results on \texttt{gemm.c} and \texttt{3mm.c}. 
In particular, the execution time of these cases on WAMR is more than 2$\times$ slower than that in Wasm3 (another \wasm interpreter), while WAMR can achieve comparable performance to Wasm3 on other normal cases.
This indicates that WAMR may not optimize the matrix multiplication operation well enough in interpreter mode.

In Issue $\#2167$, the abnormal case \texttt{flops-8.c} calculates integral of $sin(x)*cos(x)*cos(x)$ from $0$ to $PI/3$. The \ics is shown in Figure~\ref{fig:case7}. We find that the abnormal latency of WAMR occurs when handling complex arithmetic expressions in a loop, like the code at lines 249-250. We also observe this phenomenon in some other similar programs of \texttt{flops-8.c}. 
Therefore, WAMR in interpreter mode may also not have sufficient runtime optimization for such complex arithmetic expressions.
We have received confirmation for these two issues.

\subsection{RQ3: Computational Overhead}
The efficiency of \wasmdiff is important for its usability in practice. Therefore, we also evaluate the computational overhead of differential testing in \wasmdiff.
Specifically, we measure the running time of the differential testing part (abnormal case identification and performance issue location) in \wasmdiff, with different numbers of runtime settings. 
We exclude the time of performance data collection because this part of the time is determined by test case execution and should not be counted in the overhead of differential testing.
For each number of runtime settings, we perform differential testing on all possible runtime setting combinations 10 times and calculate the average running time.

The results are shown in Table~\ref{tab:overhead}. 
We can find that as the number of runtime settings grows, the computational overhead of differential testing increases steadily, but all within one second.
In our experiments, the time spent on performance data collection for a single execution of all the test cases is about two hours. It means that the computational overhead of differential testing only accounts for less than 0.01\% of the whole process.
The results indicate that \wasmdiff is highly efficient, which provides good usability for its practice.

\begin{table}[t!]
\centering
\caption{Computational overhead of differential testing under different numbers of runtime settings.}
\resizebox{\linewidth}{!}{
\begin{tabular}{l|cccccc}
 \toprule
 \textbf{\#Runtime} & 2 & 3 & 4 & 5 & 6  & 7 \\
 \midrule
 \textbf{Avg. Overhead (s)} & 0.330 & 0.476 & 0.604 & 0.735 & 0.845 & 0.966 \\
 \textbf{Std. Deviation} & 0.026 & 0.039 & 0.047 & 0.058 & 0.044 & 0.037 \\
 \bottomrule
\end{tabular}
}
\label{tab:overhead}
\end{table}

\section{Discussion}\label{sec:dis}

\subsection{Threats to Validity}
There are some threats to validity of our work, including test case selection, \wasm runtime selection, and the sufficiency of case analysis.

First, we select 123 C/C++ programs from the LLVM test suite as our test cases, which may not be very large-scale. 
However, the test cases are representative benchmark programs for performance testing and are well-suited as the source programs of \wasm. Our test cases include Polybench~\cite{polybench}, a popular benchmark that is widely used for Wasm performance evaluation in previous studies~\cite{jangda2019not, jangda2019mind, spies2021evaluation}.
Furthermore, the goal of our work is to reveal some unknown performance issues in server-side \wasm runtimes instead of finding all the performance issues (actually, it is impossible). Our evaluation has shown the effectiveness of the selected test cases in achieving our goal.

Second, we select five server-side standalone \wasm runtimes as our test targets. We select the \wasm runtimes according to their popularity and activity, thereby ensuring the representativeness of the selected runtimes.
Also, the number of runtime settings may affect the testing results, as the abnormal cases are identified based on the execution time ratio on these runtime settings. We conduct a series of experiments with different numbers of runtime settings, and we find that those abnormal cases with high \dd on the issue-related runtime can always be identified.

Third, we only report the top 10 abnormal cases in this paper, as it is inappropriate to set a threshold for abnormal case identification.
We report the 10 abnormal cases since they are with the top anomaly degree and worthy of attention. We conduct an in-depth case analysis to reveal the causes of the performance issues. We also report these issues to the developers of the corresponding \wasm runtimes to get their confirmation. The results indicate the effectiveness of our differential testing approach.

\subsection{Future Work}

In this work, we propose a novel differential testing approach to identify performance issues in server-side standalone \wasm runtimes. Based on our approach, we can collect more performance issues in existing popular standalone \wasm runtimes, then build a comprehensive benchmark suite for testing the performance issues in \wasm runtimes. This benchmark suite can facilitate future work on performance issue testing for \wasm runtimes.

Also, our work can facilitate the improvement of existing standalone \wasm runtime implementation. In the future, we aim to further improve the internal mechanisms related to the performance issues in existing \wasm runtimes. We can also design a new \wasm runtime implementation with a better optimization strategy and execution mechanism.

\section{Related Work}\label{sec:re}

\textbf{Server-side \wasm.}
WebAssembly (\wasm) is a low-level bytecode language originally designed for client-side execution in Web browsers~\cite{haas2017bringing}.
\wasm's sandboxing execution mechanism brings safety, higher-performance, lightweight, and portability natures, making it suitable for server-side applications as well~\cite{mendki2020evaluating,makitalo2021webassembly,kjorveziroski2022evaluating}.
Cloud applications built with \wasm have become increasingly popular in recent years~\cite{shillaker2020faasm,gackstatter2022pushing,eriksson2021containerizing,jain2022extending,long2020lightweight}.
For example, FASSM~\cite{shillaker2020faasm} introduces a new isolation abstraction based on \wasm for high performance serverless computing.
\wasm is also suggested to enable computational offloading in cloud environments~\cite{huang2021evaluation,nurul2021nomad,li2021wiprog,wen2020wasmachine}.
Nomad~\cite{nurul2021nomad} provides a cross-platform computational offloading and migration mechanism in Femtoclouds using \wasm.
WIPROG~\cite{li2021wiprog} proposes an edge-centric approach to IoT application programming based on \wasm.
Besides cloud environments, \wasm is also used in microcontrollers~\cite{gurdeep2019warduino,zandberg2021femto}, Trusted Execution Environments (TEEs)~\cite{menetrey2021twine} and smart contracts~\cite{zheng2020vm,wang2020wana,chen2022wasai}.

\textbf{\wasm Performance.}
High performance is an important design consideration of \wasm. \wasm attempts to provide near-native execution speed both in browsers and server-side applications~\cite{haas2017bringing, hilbig2021empirical}. 
Extensive work studies \wasm performance over the browsers~\cite{reiser2017accelerate,jangda2019not,jangda2019mind,wang2021empowering,yan2021understanding,de2022webassembly,de2021runtime}. 
Jangda \textit{et al.}~\cite{jangda2019not,jangda2019mind} build BROWSIX-Wasm to run unmodified \wasm-compiled Unix applications directly inside the browser. Then they use BROWSIX-Wasm to conduct the first large-scale evaluation of the performance of \wasm in comparison with native code. They point out a substantial performance gap between the two.
Wang~\cite{wang2021empowering} investigates how Chrome optimizes \wasm execution in comparison to JavaScript.
Yan \textit{et al.}~\cite{yan2021understanding} extend this study to more browser engines (Chrome, Firefox, and Edge). They find that JIT optimization significantly impacts JavaScript speed but has little effect on \wasm speed. Also, \wasm uses much more memory than JavaScript.
Regarding the server-side \wasm performance, there are relatively fewer studies. Spies \textit{et al.}~\cite{spies2021evaluation} conduct an evaluation of \wasm performance in non-Web environments. 
The evaluation demonstrates that \wasm is generally faster than JavaScript and can approach native code performance in some cases.
To sum up, existing studies simply compare the performance of \wasm with other codes. There is still a lack of research on how to test performance issues in \wasm runtimes.

\textbf{Differential Testing.}
Differential testing is a popular software testing technique for detecting bugs in two or more comparable systems or different implementations of the same application~\cite{mckeeman1998differential,carlson1979toward,evans2007differential,groce2007randomized}. The idea is to provide the same input to these comparable systems, and observe the inconsistency in their execution. If the results differ, it indicates that some of the systems may contain a bug.
Differential testing has been widely used to detect semantic bugs in diverse domains, such as C compilers~\cite{yang2011finding,barany2018finding,chen2016empirical,kastner2017differential}, JVM implementations~\cite{chen2016coverage,chen2019deep,zhao2022history,brennan2020jvm}, SSL/TLS implementations~\cite{brubaker2014using,chen2015guided,petsios2017nezha}, and even deep learning systems~\cite{pei2017deepxplore,guo2018dlfuzz,guo2020coverage}.
Existing differential testing approaches can be divided into two categories, unguided and guided, based on the way of input generation.
Unguided differential testing approaches generate test inputs independently without considering information from past inputs. An example is Frankencerts~\cite{brubaker2014using}, which tests for semantic violations of SSL/TLS certificate validation across multiple implementations.
Guided differential testing approaches aim to minimize the number of inputs by considering program behavior information for past inputs, making the testing process more efficient.
For example, classfuzz~\cite{chen2016coverage} is a coverage-guided fuzzing approach for differential testing of JVMs’ startup processes.
Existing differential testing approaches only focus on semantic or logic bugs in software systems. In this work, we first extend differential testing to performance issue detection, which is one of our contributions.
 \section{Conclusion}\label{sec:con}
Performance issues are critical for server-side \wasm applications, but research in this area is lacking.
In this work, we conduct the first study on performance issues in server-side standalone \wasm runtimes. We propose a novel differential testing approach \wasmdiff to identify performance issues in standalone \wasm runtimes, and we apply it to five popular real-world \wasm runtimes with 123 test cases. We further conduct a comprehensive case analysis of the top 10 identified abnormal cases, and summarize seven performance issues in four popular \wasm runtimes. All issues are confirmed by developers. The results indicate the effectiveness of \wasmdiff, which provide inspiration for future work on improving server-side \wasm runtime implementation.


\section*{Acknowledgment}
This work was supported by the Natural Science Foundation of Shanghai (No. 22ZR1407900), the Key-Area Research and Development Program of Guangdong Province (No. 2020B010165002) and the Key Program of Fundamental Research from Shenzhen Science and Technology Innovation Commission (No. JCYJ20200109113403826). It was also supported by the Research Grants Council of the Hong Kong Special Administrative Region, China (No. CUHK 14206921 of the General Research Fund).

\balance
\bibliographystyle{IEEEtran}
\bibliography{ref}

\begin{thebibliography}{10}
\providecommand{\url}[1]{#1}
\csname url@samestyle\endcsname
\providecommand{\newblock}{\relax}
\providecommand{\bibinfo}[2]{#2}
\providecommand{\BIBentrySTDinterwordspacing}{\spaceskip=0pt\relax}
\providecommand{\BIBentryALTinterwordstretchfactor}{4}
\providecommand{\BIBentryALTinterwordspacing}{\spaceskip=\fontdimen2\font plus
\BIBentryALTinterwordstretchfactor\fontdimen3\font minus
  \fontdimen4\font\relax}
\providecommand{\BIBforeignlanguage}[2]{{%
\expandafter\ifx\csname l@#1\endcsname\relax
\typeout{** WARNING: IEEEtran.bst: No hyphenation pattern has been}%
\typeout{** loaded for the language `#1'. Using the pattern for}%
\typeout{** the default language instead.}%
\else
\language=\csname l@#1\endcsname
\fi
#2}}
\providecommand{\BIBdecl}{\relax}
\BIBdecl

\bibitem{haas2017bringing}
A.~Haas, A.~Rossberg, D.~L. Schuff, B.~L. Titzer, M.~Holman, D.~Gohman,
  L.~Wagner, A.~Zakai, and J.~Bastien, ``Bringing the web up to speed with
  webassembly,'' in \emph{Proceedings of the 38th ACM SIGPLAN Conference on
  Programming Language Design and Implementation}, 2017, pp. 185--200.

\bibitem{reiser2017accelerate}
M.~Reiser and L.~Bl{\"a}ser, ``Accelerate javascript applications by
  cross-compiling to webassembly,'' in \emph{Proceedings of the 9th ACM SIGPLAN
  International Workshop on Virtual Machines and Intermediate Languages}, 2017,
  pp. 10--17.

\bibitem{wagner2017webassembly}
L.~Wagner, ``A webassembly milestone: Experimental support in multiple
  browsers,'' \emph{Mozilla Hacks (14 March 2016).}, 2017.

\bibitem{shillaker2020faasm}
S.~Shillaker and P.~Pietzuch, ``Faasm: Lightweight isolation for efficient
  stateful serverless computing,'' in \emph{2020 USENIX Annual Technical
  Conference (USENIX ATC 20)}, 2020, pp. 419--433.

\bibitem{gackstatter2022pushing}
P.~Gackstatter, P.~A. Frangoudis, and S.~Dustdar, ``Pushing serverless to the
  edge with webassembly runtimes,'' in \emph{2022 22nd IEEE International
  Symposium on Cluster, Cloud and Internet Computing (CCGrid)}.\hskip 1em plus
  0.5em minus 0.4em\relax IEEE, 2022, pp. 140--149.

\bibitem{eriksson2021containerizing}
F.~Eriksson and S.~Grunditz, ``Containerizing webassembly: Considering
  webassembly containers on iot devices as edge solution,'' 2021.

\bibitem{jain2022extending}
S.~M. Jain and S.~M. Jain, ``Extending istio with webassembly,''
  \emph{WebAssembly for Cloud: A Basic Guide for Wasm-Based Cloud Apps}, pp.
  151--160, 2022.

\bibitem{long2020lightweight}
J.~Long, H.-Y. Tai, S.-T. Hsieh, and M.~J. Yuan, ``A lightweight design for
  serverless function as a service,'' \emph{IEEE Software}, vol.~38, no.~1, pp.
  75--80, 2020.

\bibitem{dockerwasm}
\emph{Announcing Docker+Wasm Technical Preview},
  https://www.docker.com/blog/announcing-dockerwasm-technical-preview/.

\bibitem{docker2020docker}
I.~Docker, ``Docker,'' \emph{l{\i}nea].[Junio de 2017]. Disponible en:
  https://www. docker. com/what-docker}, 2020.

\bibitem{wasmedge}
\emph{WasmEdge}, https://github.com/WasmEdge/WasmEdge.

\bibitem{gurdeep2019warduino}
R.~Gurdeep~Singh and C.~Scholliers, ``Warduino: a dynamic webassembly virtual
  machine for programming microcontrollers,'' in \emph{Proceedings of the 16th
  ACM SIGPLAN International Conference on Managed Programming Languages and
  Runtimes}, 2019, pp. 27--36.

\bibitem{zandberg2021femto}
K.~Zandberg and E.~Baccelli, ``Femto-containers: Devops on microcontrollers
  with lightweight virtualization \& isolation for iot software modules,''
  \emph{arXiv preprint arXiv:2106.12553}, 2021.

\bibitem{menetrey2021twine}
J.~M{\'e}n{\'e}trey, M.~Pasin, P.~Felber, and V.~Schiavoni, ``Twine: An
  embedded trusted runtime for webassembly,'' in \emph{2021 IEEE 37th
  International Conference on Data Engineering (ICDE)}.\hskip 1em plus 0.5em
  minus 0.4em\relax IEEE, 2021, pp. 205--216.

\bibitem{zheng2020vm}
S.~Zheng, H.~Wang, L.~Wu, G.~Huang, and X.~Liu, ``Vm matters: A comparison of
  wasm vms and evms in the performance of blockchain smart contracts,''
  \emph{arXiv preprint arXiv:2012.01032}, 2020.

\bibitem{wang2020wana}
D.~Wang, B.~Jiang, and W.~Chan, ``Wana: Symbolic execution of wasm bytecode for
  cross-platform smart contract vulnerability detection,'' \emph{arXiv preprint
  arXiv:2007.15510}, 2020.

\bibitem{chen2022wasai}
W.~Chen, Z.~Sun, H.~Wang, X.~Luo, H.~Cai, and L.~Wu, ``Wasai: uncovering
  vulnerabilities in wasm smart contracts,'' in \emph{Proceedings of the 31st
  ACM SIGSOFT International Symposium on Software Testing and Analysis}, 2022,
  pp. 703--715.

\bibitem{jangda2019not}
A.~Jangda, B.~Powers, E.~D. Berger, and A.~Guha, ``Not so fast: Analyzing the
  performance of $\{$WebAssembly$\}$ vs. native code,'' in \emph{2019 USENIX
  Annual Technical Conference (USENIX ATC 19)}, 2019, pp. 107--120.

\bibitem{jangda2019mind}
A.~Jangda, B.~Powers, A.~Guha, and E.~Berger, ``Mind the gap: Analyzing the
  performance of webassembly vs. native code,'' \emph{arXiv preprint
  arXiv:1901.09056}, 2019.

\bibitem{wang2021empowering}
W.~Wang, ``Empowering web applications with webassembly: Are we there yet?'' in
  \emph{2021 36th IEEE/ACM International Conference on Automated Software
  Engineering (ASE)}.\hskip 1em plus 0.5em minus 0.4em\relax IEEE, 2021, pp.
  1301--1305.

\bibitem{yan2021understanding}
Y.~Yan, T.~Tu, L.~Zhao, Y.~Zhou, and W.~Wang, ``Understanding the performance
  of webassembly applications,'' in \emph{Proceedings of the 21st ACM Internet
  Measurement Conference}, 2021, pp. 533--549.

\bibitem{de2022webassembly}
J.~De~Macedo, R.~Abreu, R.~Pereira, and J.~Saraiva, ``Webassembly versus
  javascript: Energy and runtime performance,'' in \emph{2022 International
  Conference on ICT for Sustainability (ICT4S)}.\hskip 1em plus 0.5em minus
  0.4em\relax IEEE, 2022, pp. 24--34.

\bibitem{de2021runtime}
------, ``On the runtime and energy performance of webassembly: Is webassembly
  superior to javascript yet?'' in \emph{2021 36th IEEE/ACM International
  Conference on Automated Software Engineering Workshops (ASEW)}.\hskip 1em
  plus 0.5em minus 0.4em\relax IEEE, 2021, pp. 255--262.

\bibitem{spies2021evaluation}
B.~Spies and M.~Mock, ``An evaluation of webassembly in non-web environments,''
  in \emph{2021 XLVII Latin American Computing Conference (CLEI)}.\hskip 1em
  plus 0.5em minus 0.4em\relax IEEE, 2021, pp. 1--10.

\bibitem{awesome}
\emph{Awesome WebAssembly Runtimes},
  https://github.com/appcypher/awesome-wasm-runtimes.

\bibitem{mckeeman1998differential}
W.~M. McKeeman, ``Differential testing for software,'' \emph{Digital Technical
  Journal}, vol.~10, no.~1, pp. 100--107, 1998.

\bibitem{carlson1979toward}
J.~S. Carlson and K.~H. Wiedl, ``Toward a differential testing approach:
  Testing-the-limits employing the raven matrices,'' \emph{Intelligence},
  vol.~3, no.~4, pp. 323--344, 1979.

\bibitem{evans2007differential}
R.~B. Evans and A.~Savoia, ``Differential testing: a new approach to change
  detection,'' in \emph{The 6th Joint Meeting on European software engineering
  conference and the ACM SIGSOFT Symposium on the Foundations of Software
  Engineering: Companion Papers}, 2007, pp. 549--552.

\bibitem{groce2007randomized}
A.~Groce, G.~Holzmann, and R.~Joshi, ``Randomized differential testing as a
  prelude to formal verification,'' in \emph{29th International Conference on
  Software Engineering (ICSE'07)}.\hskip 1em plus 0.5em minus 0.4em\relax IEEE,
  2007, pp. 621--631.

\bibitem{wasmer}
\emph{Wasmer}, https://github.com/wasmerio/wasmer.

\bibitem{wasmtime}
\emph{Wasmtime}, https://github.com/bytecodealliance/wasmtime.

\bibitem{wasm3}
\emph{Wasm3}, https://github.com/wasm3/wasm3.

\bibitem{wamr}
\emph{WebAssembly Micro Runtime},
  https://github.com/bytecodealliance/wasm-micro-runtime.

\bibitem{llvmtest}
\emph{test-suite Guide}, https://llvm.org/docs/TestSuiteGuide.html.

\bibitem{zakai2011emscripten}
A.~Zakai, ``Emscripten: an llvm-to-javascript compiler,'' in \emph{Proceedings
  of the ACM international conference companion on Object oriented programming
  systems languages and applications companion}, 2011, pp. 301--312.

\bibitem{hilbig2021empirical}
A.~Hilbig, D.~Lehmann, and M.~Pradel, ``An empirical study of real-world
  webassembly binaries: Security, languages, use cases,'' in \emph{Proceedings
  of the Web Conference 2021}, 2021, pp. 2696--2708.

\bibitem{clark2019standardizing}
L.~Clark, ``Standardizing wasi: A system interface to run webassembly outside
  the web,'' \emph{Mozilla Hacks--the Web developer blog}, 2019.

\bibitem{cncf}
\emph{Cloud Native Computing Foundation}, https://www.cncf.io/.

\bibitem{bytecodealliance}
\emph{Bytecode Alliance}, https://bytecodealliance.org/.

\bibitem{microservice}
\emph{Secure \& lightweight microservice with a database backend},
  https://github.com/second-state/microservice-rust-mysql.

\bibitem{ab}
\emph{ab - Apache HTTP server benchmarking tool},
  https://httpd.apache.org/docs/2.4/programs/ab.html.

\bibitem{perf}
\emph{perf: Linux profiling with performance counters},
  https://perf.wiki.kernel.org/index.php/Main\_Page.

\bibitem{polybench}
\emph{PolyBench/C - the Polyhedral Benchmark suite},
  https://web.cse.ohio-state.edu/~pouchet.2/software/polybench/.

\bibitem{cranelift}
\emph{Cranelift Code Generator},
  https://github.com/bytecodealliance/wasmtime/\\tree/main/cranelift.

\bibitem{mendki2020evaluating}
P.~Mendki, ``Evaluating webassembly enabled serverless approach for edge
  computing,'' in \emph{2020 IEEE Cloud Summit}.\hskip 1em plus 0.5em minus
  0.4em\relax IEEE, 2020, pp. 161--166.

\bibitem{makitalo2021webassembly}
N.~M{\"a}kitalo, T.~Mikkonen, C.~Pautasso, V.~Bankowski, P.~Daubaris,
  R.~Mikkola, and O.~Beletski, ``Webassembly modules as lightweight containers
  for liquid iot applications,'' in \emph{Web Engineering: 21st International
  Conference, ICWE 2021, Biarritz, France, May 18--21, 2021,
  Proceedings}.\hskip 1em plus 0.5em minus 0.4em\relax Springer, 2021, pp.
  328--336.

\bibitem{kjorveziroski2022evaluating}
V.~Kjorveziroski, S.~Filiposka, and A.~Mishev, ``Evaluating webassembly for
  orchestrated deployment of serverless functions,'' in \emph{2022 30th
  Telecommunications Forum (TELFOR)}.\hskip 1em plus 0.5em minus 0.4em\relax
  IEEE, 2022, pp. 1--4.

\bibitem{huang2021evaluation}
W.~Huang and M.~Paradies, ``An evaluation of webassembly and ebpf as offloading
  mechanisms in the context of computational storage,'' \emph{arXiv preprint
  arXiv:2111.01947}, 2021.

\bibitem{nurul2021nomad}
M.~Nurul-Hoque and K.~A. Harras, ``Nomad: Cross-platform computational
  offloading and migration in femtoclouds using webassembly,'' in \emph{2021
  IEEE International Conference on Cloud Engineering (IC2E)}.\hskip 1em plus
  0.5em minus 0.4em\relax IEEE, 2021, pp. 168--178.

\bibitem{li2021wiprog}
B.~Li, W.~Dong, and Y.~Gao, ``Wiprog: A webassembly-based approach to
  integrated iot programming,'' in \emph{IEEE INFOCOM 2021-IEEE Conference on
  Computer Communications}.\hskip 1em plus 0.5em minus 0.4em\relax IEEE, 2021,
  pp. 1--10.

\bibitem{wen2020wasmachine}
E.~Wen and G.~Weber, ``Wasmachine: Bring iot up to speed with a webassembly
  os,'' in \emph{2020 IEEE International Conference on Pervasive Computing and
  Communications Workshops (PerCom Workshops)}.\hskip 1em plus 0.5em minus
  0.4em\relax IEEE, 2020, pp. 1--4.

\bibitem{yang2011finding}
X.~Yang, Y.~Chen, E.~Eide, and J.~Regehr, ``Finding and understanding bugs in c
  compilers,'' in \emph{Proceedings of the 32nd ACM SIGPLAN conference on
  Programming language design and implementation}, 2011, pp. 283--294.

\bibitem{barany2018finding}
G.~Barany, ``Finding missed compiler optimizations by differential testing,''
  in \emph{Proceedings of the 27th international conference on compiler
  construction}, 2018, pp. 82--92.

\bibitem{chen2016empirical}
J.~Chen, W.~Hu, D.~Hao, Y.~Xiong, H.~Zhang, L.~Zhang, and B.~Xie, ``An
  empirical comparison of compiler testing techniques,'' in \emph{Proceedings
  of the 38th International Conference on Software Engineering}, 2016, pp.
  180--190.

\bibitem{kastner2017differential}
C.~K{\"a}stner, ``Differential testing for variational analyses: Experience
  from developing kconfigreader,'' \emph{arXiv preprint arXiv:1706.09357},
  2017.

\bibitem{chen2016coverage}
Y.~Chen, T.~Su, C.~Sun, Z.~Su, and J.~Zhao, ``Coverage-directed differential
  testing of jvm implementations,'' in \emph{proceedings of the 37th ACM
  SIGPLAN Conference on Programming Language Design and Implementation}, 2016,
  pp. 85--99.

\bibitem{chen2019deep}
Y.~Chen, T.~Su, and Z.~Su, ``Deep differential testing of jvm
  implementations,'' in \emph{2019 IEEE/ACM 41st International Conference on
  Software Engineering (ICSE)}.\hskip 1em plus 0.5em minus 0.4em\relax IEEE,
  2019, pp. 1257--1268.

\bibitem{zhao2022history}
Y.~Zhao, Z.~Wang, J.~Chen, M.~Liu, M.~Wu, Y.~Zhang, and L.~Zhang,
  ``History-driven test program synthesis for jvm testing,'' in
  \emph{Proceedings of the 44th International Conference on Software
  Engineering}, 2022, pp. 1133--1144.

\bibitem{brennan2020jvm}
T.~Brennan, S.~Saha, and T.~Bultan, ``Jvm fuzzing for jit-induced side-channel
  detection,'' in \emph{Proceedings of the ACM/IEEE 42nd International
  Conference on Software Engineering}, 2020, pp. 1011--1023.

\bibitem{brubaker2014using}
C.~Brubaker, S.~Jana, B.~Ray, S.~Khurshid, and V.~Shmatikov, ``Using
  frankencerts for automated adversarial testing of certificate validation in
  ssl/tls implementations,'' in \emph{2014 IEEE Symposium on Security and
  Privacy}.\hskip 1em plus 0.5em minus 0.4em\relax IEEE, 2014, pp. 114--129.

\bibitem{chen2015guided}
Y.~Chen and Z.~Su, ``Guided differential testing of certificate validation in
  ssl/tls implementations,'' in \emph{Proceedings of the 2015 10th Joint
  Meeting on Foundations of Software Engineering}, 2015, pp. 793--804.

\bibitem{petsios2017nezha}
T.~Petsios, A.~Tang, S.~Stolfo, A.~D. Keromytis, and S.~Jana, ``Nezha:
  Efficient domain-independent differential testing,'' in \emph{2017 IEEE
  Symposium on security and privacy (SP)}.\hskip 1em plus 0.5em minus
  0.4em\relax IEEE, 2017, pp. 615--632.

\bibitem{pei2017deepxplore}
K.~Pei, Y.~Cao, J.~Yang, and S.~Jana, ``Deepxplore: Automated whitebox testing
  of deep learning systems,'' in \emph{proceedings of the 26th Symposium on
  Operating Systems Principles}, 2017, pp. 1--18.

\bibitem{guo2018dlfuzz}
J.~Guo, Y.~Jiang, Y.~Zhao, Q.~Chen, and J.~Sun, ``Dlfuzz: Differential fuzzing
  testing of deep learning systems,'' in \emph{Proceedings of the 2018 26th ACM
  Joint Meeting on European Software Engineering Conference and Symposium on
  the Foundations of Software Engineering}, 2018, pp. 739--743.

\bibitem{guo2020coverage}
J.~Guo, Y.~Zhao, H.~Song, and Y.~Jiang, ``Coverage guided differential
  adversarial testing of deep learning systems,'' \emph{IEEE Transactions on
  Network Science and Engineering}, vol.~8, no.~2, pp. 933--942, 2020.

\end{thebibliography}


\end{document}